\documentclass[aps,11pt,preprintnumbers,nofootinbib,floatfix]{revtex4}

\usepackage{graphicx}
\usepackage{epsfig}
\usepackage{dcolumn}
\usepackage{subfigure}
\usepackage{multirow}
\usepackage{amsmath}
\usepackage{amssymb}

\begin{document}

\title{Charged AdS black holes in Gauss-Bonnet gravity and nonlinear electrodynamics}


\author{Seungjoon Hyun$^{1}$\footnote{email: sjhyun@yonsei.ac.kr} and Cao H. Nam$^{2,1}$\footnote{email: hncao@yonsei.ac.kr}}

\affiliation{\small $^{1}$Department of Physics,
College of Science, Yonsei University, Seoul 120-749, Korea\\
$^{2}$Institute of Research and Development, Duy Tan 
University, Da Nang 550000, Vietnam}
\date{\today}

\begin{abstract}%
New five-dimensional charged AdS black hole solutions are found in Einstein-Gauss-Bonnet gravity and the nonlinear electrodynamics. These solutions include regular black holes as well as extremal black holes. The first law of the black hole thermodynamics is confirmed in the extended phase space where the cosmological constant is treated as the pressure. The first and second order phase transitions are investigated by observing the behavior of the heat capacity at constant pressure and the Gibbs free energy. 
In addition, the equation of state for the black holes and their $P-V$ criticality are studied. 
Finally, the critical exponents are found to  be the same as those of the Van der Waals fluid.
\end{abstract}

\maketitle

\section{Introduction}

Black holes are the solutions of Einstein's equation and one of the most important and fascinating objects in physics. They themselves possess a lot of  interesting physical properties. As found by Bekenstein and Hawking, the black holes are thermodynamic objects with the temperature and entropy determined by the surface gravity at the event horizon and by the area of the event horizon, respectively \cite{Bekenstein72,Bekenstein74,Bekenstein73,Carter,Hawking75}. Since the pioneering work of Bekenstein and Hawking, the thermodynamics and the phase structure of black holes have been studied extensively \cite{Hut1977,Davies1978,Mazur1980,Pavon1991}. In particular, the black hole thermodynamics in the asymptotically anti-de Sitter (AdS) spacetime has attracted a lot of attention. In $1983$, Hawking and Page discovered a phase transition between Schwarzschild AdS black hole and thermal AdS space, so-called, the Hawking-Page transition \cite{Hawking1983}. Interestingly, according to the AdS/CFT correspondence, the Hawking-Page transition is interpreted as the confinement/deconfinement phase transition  in the dual conformal field theory at the boundary \cite{Maldacena,Witten,Gubser,Aharony}. Chamblin \textit{et al.} found that charged AdS black holes  can undergo a first-order phase transition  which is very similar to that of the liquid/gas phase system \cite{Chamblin199a,Chamblin199b}. 

Recently, there have been attempts to treat the cosmological constant  as the thermodynamic pressure $P$ whose conjugate variable is the thermodynamic volume $V$. In this extension, the first law of black hole thermodynamics is modified with including a new term $VdP$ \cite{Wang2006,Kastor2009,Kastor2010,Dolan2011,Dolan2011b,Teo2017} and the black hole mass is naturally interpreted as the enthalpy rather than the internal energy. Accordingly, the study of the thermodynamics and phase transition for the AdS black holes has been generalized to the extended phase space. $P-V$ criticality of charged AdS black holes was found by Kubiz\v{n}\'{a}k and Mann, at which the critical exponents were obtained and shown to coincide with those of the Van der Waals system \cite{Mann2012}. Following this remarkable work, $P-V$ criticality has been studied in various black holes \cite{Gunasekaran2012,Chabab2012,Hendi2012,Cai2013,Liu2013,Hendi2013,MoLiu2014,Liu2014,Li2014,Zhao2014,
Dehghani2014,Hennigar2015,Hendi2015d,Xu2015,Talezadeh2016,Sadeghi2016a,Liang2016,Fernando2016,Fan2016,Sadeghi2016b,Hansen2017,Majhi2017,Talezadeh2017, Upadhyay2017,Dayyani2018,Nam2018a,Pradhan2018,Nam2018d}. In addition, many interesting phenomena of the AdS black hole thermodynamics have been found, such as multiply reentrant phase transitions \cite{Gunasekaran2012,Altamirano2013,Frassino2014,Henninger2015}, the heat engine efficiency of black holes \cite{Johson2014,Belhaj2015,Setare2015,Johson2016a,Johson2016b,Zhang2016,Bhamidipati2017,Hennigar17,Mo2017,
Hendi2018a,Nam2019}, or the Joule-Thomson expansion of black holes \cite{Aydner2017a,Aydner2017b,Xu2018,Chabab2018,Mo-Li2018,Lan2018}.

Einstein-Gauss-Bonnet gravity is a natural generalization of Einstein gravity in $(4+1)$ dimensions or higher in the gravity model which includes the higher order curvature corrections. This theory corresponds to first three terms of Lovelock gravity, which consists of  the cosmological constant, Einstein-Hilbert action and the Gauss-Bonnet term \cite{Lovelock1971}. The Gauss-Bonnet term is also naturally obtained from the low-energy limit of heterotic string theory \cite{Zwiebach1985,Witten1986,Gross1987,Tseytlin1987,Bento1996}. Noble features of the theory including the Gauss-Bonnet term are that the theory is ghost-free and the corresponding field equations do not contain more than second order derivatives of the metric \cite{Zumino1986,Myers1987,Callan1989,Cho2002}. In Einstein-Gauss-Bonnet gravity, the black hole solutions and their interesting properties have been studied in the literature \cite{Boulware1985,Wiltshire1988,Cai2002,Guo2004,Alexeyev2004,Troncoso2007,Charmousis2009,Panah2010,Yang2013,Hendi2014,
Ferrari2015,Panah2016}.

One of the mysterious properties of black holes is that they have a curvature singularity in their interior \cite{Hawking1973}. This curvature singularity is inevitable in General Relativity under some physically reasonable conditions because of the famous singularity theorem proved by Hawking and Penrose \cite{Hawking1973}. The presence of the curvature singularity may be regarded as signaling the breakdown of General Relativity and would be removed by quantum gravity, which is not known yet. Since lots of efforts have been dedicated to understand how to avoid the curvature singularity at the semi-classical level. The first example of such singularity-free, or so-called regular, black hole was found by Bardeen \cite{Bardeen}, which was realized later as a gravitational collapse of some magnetic monopole in the framework of the nonlinear electrodynamics by Ayon-Beato and Garcia \cite{Ayon-Beato2000}. 

There have been a lot of the regular black hole solutions obtained with the nonlinear electrodynamics. In Refs. \cite{Ayon-Beato1999a,Ayon-Beato1999b,Bronnikov2001}, they proposed the regular black holes which behave asymptotically like the RN black hole but do not satisfy the weak energy condition (WEC). The Bardeen regular black hole \cite{Bardeen,Ayon-Beato2000} satisfies WEC, but its asymptotic behavior does not look like the one of  the RN black hole. The regular black holes, which satisfy WEC and behave asymptotically like the RN black hole, were obtained by authors in Refs. \cite{Ayon-Beato98,Dymnikova2004,Schee2015,Gan2016}. In addition, the regular black hole solutions were found from the modification of the Ayon-Beato and Garcia solution \cite{Burinskii2002}, in the lower dimensional spacetime \cite{Cataldo2000}, in the presence of some dark energy candidates \cite{Nam2018b,Nam2018c}. There exist other regular black hole solutions such as
rotating regular black hole \cite{Toshmatov} and radiating Kerr-like regular black hole \cite{Ghosh}. Extremal limit of the regular charged black holes was investigated in Ref. \cite{Matyjasek2004}. Beyond Einstein gravity, the regular black hole solutions were found in Einstein-Gauss-Bonnet gravity \cite{Singh2018,SGGosh2019}, in the quadratic gravity \cite{Berej2006}, in the $f(R)$ gravity \cite{Nojiri2017}, and in the $f(T)$ gravity \cite{Junior2015}. 

Inspired by these works,  we obtain new charged AdS
black hole solutions with the nonlinear electrodynamics in the framework of Einstein-Gauss-Bonnet gravity. The solutions contain not only  extremal charged black holes but also  regular black holes without curvature singularity. This means there are a variety of scenarios for the fate of these black holes after Hawking radiation. We investigate  the thermodynamics and the phase transitions of these black holes in the extended phase space including the cosmological constant as the pressure. We find that the solutions exhibit the Hawking-Page transition through the study of Gibbs free energy. We also find that the solutions have the liquid/gas-like phase transition below the critical temperature.

This paper is organized as follows. In Sec. \ref{bhsol}, we introduce the action describing Einstein-Gauss-Bonnet gravity coupled to a nonlinear electromagnetic field in five-dimensional spacetime with negative cosmological constant. Then, we find static and spherically symmetric black hole solutions carrying the electric charge. In Sec. \ref{prop}, we study the curvature singularity and find the regular black hole solutions. We investigate the thermodynamics and phase structure of the black holes in Sec. \ref{therm}. We find the first and second order phase transitions. We study $P-V$ criticality of the black holes in Sec. \ref{PV-crt}. We also obtain the critical exponents describing the behavior of the thermodynamic quantities of the black holes near the critical point. Finally, our conclusion is drawn in Sec. \ref{conclu}. Note that, in this paper, we use units in $G_N=\hbar=c=k_B=1$ and the signature of the metric $(-,+,+,+,+)$.

\section{\label{bhsol} The black hole solution}
We consider five-dimensional Einstein-Gauss-Bonnet gravity coupled to a nonlinear
electromagnetic field along with negative cosmological constant. The system is described by the action
\begin{equation}
S=\int
d^5x\sqrt{-g}\left[\frac{1}{16\pi}\left(R-2\Lambda+\alpha \mathcal{L}_{GB}\right)-\frac{1}{4\pi}\mathcal{L}(F)\right],\label{EGB-nlED-adS}
\end{equation}
where
$\Lambda$ is the negative cosmological constant related to
the AdS curvature radius $l$ as
\begin{equation}
  \Lambda=-\frac{6}{l^2},
\end{equation}
and $\mathcal{L}_{GB}$ is the Gauss-Bonnet term given by
\begin{equation}
  \mathcal{L}_{GB}=R^2-4R_{\mu\nu}R^{\mu\nu}+R_{\mu\nu\rho\lambda}R^{\mu\nu\rho\lambda},
\end{equation}
$\alpha$ denotes the Gauss-Bonnet coupling constant\footnote{The Gauss-Bonnet term appears in the
heterotic string theory at which $\alpha$ is regarded as the inverse string tension \cite{Zwiebach1985,Witten1986,Gross1987,Tseytlin1987,Bento1996}. Motivated from this, we restrict to the case $\alpha\geq0$.}, and $\mathcal{L}(F)$ is the function of the invariant
$F\equiv F_{\mu\nu}F^{\mu\nu}/4$ where $F_{\mu\nu}=\partial_\mu
A_\nu-\partial_\nu A_\mu$ is the field strength of the
electromagnetic field $A_{\mu}$. 

By varying the above action, we derive the equations of motion as
\begin{eqnarray}
G^\nu_\mu+\alpha H^\nu_\mu-\frac{6}{l^2}\delta^\nu_\mu&=&2\left[\frac{\partial\mathcal{L}(F)}{\partial
F}F_{\mu\rho}F^{\nu\rho}-\delta^\nu_\mu\mathcal{L}(F)\right],\label{Eeq}\\
\nabla_\mu\left(\frac{\partial\mathcal{L}(F)}{\partial
F}F^{\nu\mu}\right)&=&0,\label{NLEeq}
\end{eqnarray}
with
\begin{equation}
  H^\nu_\mu\equiv 2\left(RR^\nu_\mu-2R_{\mu\sigma}R^{\sigma\nu}-2R^{\sigma\rho}R^\nu_{\sigma\mu\rho}+R^{\sigma\rho\lambda}_\mu R^\nu_{\sigma\rho\lambda}\right)-\frac{1}{2}\delta^\nu_\mu\mathcal{L}_{GB}.
\end{equation}

It is convenient to introduce a two-form field strength $P^{\mu\nu}$ as
\begin{equation}
 P^{\mu\nu}\equiv \frac{\partial\mathcal{L}(F)}{\partial
F}F^{\mu\nu},
\end{equation}
whose $(tr)$ component is related to the conjugate momentum of the gauge field $A_{\mu}$. The square invariant of the field strength $P_{\mu\nu}$
satisfies
 \begin{equation}
P\equiv
\frac{1}{4}P_{\mu\nu}P^{\mu\nu}=\left(\frac{\partial\mathcal{L}(F)}{\partial
F}\right)^2F.
\end{equation}
It is also convenient to introduce $\mathcal{H}(P)$ which is related to the Lagrangian $\mathcal{L}(F)$ of the gauge field by 
a Legendre
transformation as \cite{Salazar87}
\begin{equation}
\mathcal{H}(P)=2\frac{\partial\mathcal{L}(F)}{\partial
F}F-\mathcal{L}(F). \end{equation}
and satisfies the following relations:
\begin{equation}
\mathcal{L}=2\frac{\partial\mathcal{H}}{\partial P}P-\mathcal{H}, \qquad \frac{\partial\mathcal{L}}{\partial
F}=\left(\frac{\partial\mathcal{H}}{\partial P}\right)^{-1}.\end{equation}
Then the equations of motion can be written in terms of $P_{\mu\nu}$ and  $\mathcal{H}$  as
\begin{eqnarray}
G^\nu_\mu+\alpha H^\nu_\mu-\frac{6}{l^2}\delta^\nu_\mu&=&2\left[\frac{\partial\mathcal{H}}{\partial
P}P_{\mu\rho}P^{\nu\rho}-\delta^\nu_\mu\left(2\frac{\partial\mathcal{H}}{\partial P}P-\mathcal{H}\right)\right],\label{Eeq1}\\
\nabla_\mu P^{\nu\mu}&=&0.\label{Meq}
\end{eqnarray}

As we are motivated to find black hole solutions which contain  the regular black hole solution in some special limit, we would like to consider the nonlinear electrodynamics given explicitly as
\begin{eqnarray}\label{non-eltr-term}
 \mathcal{H}(P)&=&Pe^{-k_{0}\left(-P\right)^{\frac{1}{3}}},
\end{eqnarray}
where $k_{0}$
is a coupling constant of the nonlinear electrodynamics.
The Lagrangian of the nonlinear electrodynamics corresponding to (\ref{non-eltr-term}) is given in the expansion form as 
\begin{eqnarray}\label{L-NL}
\mathcal{L}(F)=F\left[1+k_{0}\left(-F\right)^{\frac{1}{3}}+\frac{23}{18}k_{0}^{2}\left(-F\right)^{\frac{2}{3}}+\mathcal{O}(k_{0}^3)\right].
\end{eqnarray}
One may consider
the Lagrangian in (\ref{L-NL}) as the effective Lagrangian of the $U(1)$ gauge field, where the first- and higher-order terms in the nonlinear coupling constant $k_{0}$ are the higher order derivative corrections of the $U(1)$ gauge field originating from the integration of UV energy modes. The usual Maxwell electrodynamics is restored in the  $k_{0}\rightarrow 0$ limit. 

If the Gauss-Bonnet coupling constant is in the region $0\leq\alpha \leq\frac{l^2}{8}$, it admits AdS vacuum solution of the form
\begin{eqnarray}
ds^2&=&-\left(\frac{r^2}{l^2_{\textrm{eff}}}+1\right)dt^2+\frac{dr^2}{\frac{r^2}{l^2_{\textrm{eff}}}+1}+r^2d\Omega^2_3,
\label{vac-sol}
\end{eqnarray}
where the effective curvature radius $l_{\textrm{eff}}$ is given by
$
l^2_{\textrm{eff}}=\frac{l^2}{2}\left(1\pm\sqrt{1-\frac{8\alpha}{l^2}}\right).
$
Therefore,  for $\alpha< \frac{l^{2}}{8}$, we have two AdS vacuum solutions with two different effective curvature radii, while,  for $\alpha= \frac{l^{2}}{8}$, the theory has a unique AdS vacuum with the effective curvature radius, $l_{\textrm{eff}}=\frac{l}{\sqrt{2}}$. From now on, we consider the theory with $0\leq\alpha \leq\frac{l^2}{8}$ and choose  the vacuum solution with the effective curvature radius 
\begin{equation}
l^2_{\textrm{eff}}=\frac{l^2}{2}\left(1+\sqrt{1-\frac{8\alpha}{l^2}}\right)
\end{equation}
which has smooth $\alpha\rightarrow 0$ limit.

Now we would like to find a static and spherically symmetric black hole solution of the
total mass $M$ and the total electric charge $Q$, given by the following ansatz:
\begin{eqnarray}
ds^2&=&-f(r)dt^2+f(r)^{-1}dr^2+r^2d\Omega^2_3,\nonumber\\
P_{tr}&=&\phi(r).\label{charge-P}
\end{eqnarray}
The total electric-charge $Q$ is defined as
\begin{equation}
  Q=\frac{1}{4\pi}\int_{S^\infty_3}*\frac{\partial\mathcal{L}(F)}{\partial
F}\textit{\textbf{F}}=\frac{1}{4\pi}\int_{S^\infty_3}*\textit{\textbf{P}},\label{el-char}
\end{equation}
where $\textit{\textbf{F}}=\frac{1}{2}F_{\mu\nu}dx^\mu\wedge dx^\nu$ and $\textit{\textbf{P}}=\frac{1}{2}P_{\mu\nu}dx^\mu\wedge dx^\nu$. From Eqs. (\ref{Meq}) and (\ref{el-char}), with ansatz (\ref{charge-P}), we obtain
\begin{equation}
\phi(r)=\frac{q}{r^3},\ \ \qquad \ \
P=-\frac{q^2}{2r^6},
\end{equation}
where, for convenience, we introduce the 'reduced' charge $q$ related to the total charge $Q$ by $q=\frac{4\pi}{S_3}Q$ in which $S_3=2\pi^2$ is the volume of the unit $3$-sphere. 
The electrostatic potential $A_t(r)$ of the black hole is obtained as
\begin{eqnarray}\label{potential}
A_t(r)=-\int F_{tr}dr=-\int
\frac{\partial\mathcal{H}}{\partial P}P_{tr}dr
=\frac{q}{3}\left(\frac{1}{2r^2}-\frac{1}{k}\right)e^{-\frac{k}{r^2}}+\frac{q}{3k},
\end{eqnarray}
with
$k\equiv k_{0}\left(\frac{q^2}{2}\right)^{1/3}$. The integration constant is fixed to satisfy $A_t(r)\rightarrow \frac{q}{2r^{2}}$ as $r\rightarrow\infty$. 

The $(tt)$ component of Eq. (\ref{Eeq}) reads
\begin{eqnarray}\label{mrf}
\left[4\alpha f'(r)-2r\right]\left[f(r)-1\right]-r^2f'(r)+\frac{4r^3}{l^2}=\frac{2q^2}{3r^3}e^{-\frac{k}{r^2}}.
\end{eqnarray}
It is straightforward to confirm that all the other components of  the equation (\ref{Eeq}) are satisfied if $(tt)$ component equation (\ref{mrf}) holds, as presented in the appendix.  
By integrating the equation, we obtain the black hole solution as
\begin{eqnarray}
f(r)=1+\frac{r^2}{4\alpha}\left(1-\sqrt{1-\frac{8\alpha}{l^2}+\frac{8\alpha }{r^4}\left(m+\frac{q^{2}}{3k}(e^{-\frac{k}{r^2}}-1)\right)}\right),\label{fr-funct}
\end{eqnarray}
where the 'reduced' mass $m$ is the integration constant.
The ADM mass $M$ of the geometry can be obtained by following the expression given in  Ref. \cite{Rodrigo2011}  for the black holes in Einstein-Gauss-Bonnet gravity coupled to nonlinear electrodynamics  and  is given by 
\begin{eqnarray}
M=\frac{S_3}{16\pi}\left[r^3f'-12\alpha rf'(f-1)-\frac{3l^2_{\text{eff}}}{4}\left(1-\frac{12\alpha}{l^2_{\text{eff}}}\right)r^3f'\int^1_0 dt\left(\frac{1-f}{r^2}+\frac{t^2}{l^2_{\text{eff}}}\right)\right]_{r\rightarrow\infty}.
\end{eqnarray}
Since the asymptotic expansion of the metric function $f(r)$ is given by
\begin{equation}
f(r)=1+\frac{r^2}{l^2_{\text{eff}}}-\frac{1}{1-\frac{4\alpha}{l^2_{\text{eff}}}}\frac{m}{r^2}+\frac{q^2}{3\left(1-\frac{4\alpha}{l^2_{\text{eff}}}\right)}\frac{1}{r^4}+\mathcal{O}(1/r^6),
\end{equation}
 the ADM mass $M$ of the black hole solution is related to   the 'reduced' mass $m$ as
\begin{equation}
M=\frac{3S_3}{16\pi}m.
\end{equation}
One may note that, if the reduced mass is in the region $m<\frac{q^2}{3k}$, the metric has branch point  singularity at $r=r_{b}$, which is determined as
\begin{eqnarray}
1-\frac{8\alpha}{l^2}+\frac{8\alpha }{r_b^4}\left(m+\frac{q^{2}}{3k}(e^{-\frac{k}{r_b^2}}-1)\right)=0.
\end{eqnarray}
Therefore the reduced mass must satisfy the inequality,
$m\geq\frac{q^2}{3k}.$ As will be shown in the next section, this lower bound is a special point where the curvature singularity disappears.

Before we describe the solution in detail, we present various limits of the solution, which correspond to the known black hole solutions in the literatures. 
\begin{itemize}
\item{\bf $5D$ Schwarzschild-AdS GB black holes} \\In the limit $q\rightarrow0$, the metric function is given by
\begin{equation}
  f(r)=1+\frac{r^2}{4\alpha}\left(1-\sqrt{1-\frac{8\alpha}{l^2}+\frac{8\alpha m}{r^4}}\right).
\end{equation}
which corresponds to the metric of $5D$ Schwarzschild-AdS GB black hole \cite{Cai2002}. 
\item{\bf $5D$ charged AdS black holes with the GB term and Maxwell electrodynamics}\\ In the limit $k\rightarrow 0$, the Lagrangian for the gauge field becomes the usual Maxwell Lagrangian. In this limit, the solution becomes of the form,
\begin{eqnarray}
f(r)=1+\frac{r^2}{4\alpha}\left(1-\sqrt{1-\frac{8\alpha}{l^2}+\frac{8\alpha m}{r^4}-\frac{8\alpha q^{2}}{3r^6}}\right),
\end{eqnarray}
which describes the charged AdS black hole solution in Einstein-Gauss-Bonnet gravity.
If we take additional limit, $l\rightarrow\infty$, it becomes the Wiltshire black hole solution \cite{Wiltshire1988}, in which the GB term plays the role of the effective cosmological constant. On the other hand, under the limit, $l\rightarrow\infty$ and $q\rightarrow0$, it reduces to the Boulware-Deser black hole solution \cite{Boulware1985}.
\item{\bf $5D$ charged AdS black holes with higher derivative electrodynamics}\\ If we take the $\alpha\rightarrow0$ limit, we obtain another new charged black hole solution of Einstein gravity with higher derivative $U(1)$ gauge field as 
\begin{eqnarray}
f(r)=\frac{r^2}{l^{2}}+1-\frac{m}{r^2}-\frac{q^2}{3kr^2}\Big(e^{-\frac{k}{r^2}}-1\Big).
\end{eqnarray}
In this solution, the lower bound of the reduced  mass is also given by $m = q^2/3k$, where it becomes the regular black hole solution.
Since all the thermodynamic properties of this solution can be trivially obtained by taking $\alpha\rightarrow 0$ limit, we will present the thermodynamic properties for the general $\alpha$ case only.
\end{itemize}

\section{Properties of black hole Solutions}\label{prop}

\begin{figure}[t]
 \centering
\begin{tabular}{cc}
\includegraphics[width=0.45 \textwidth]{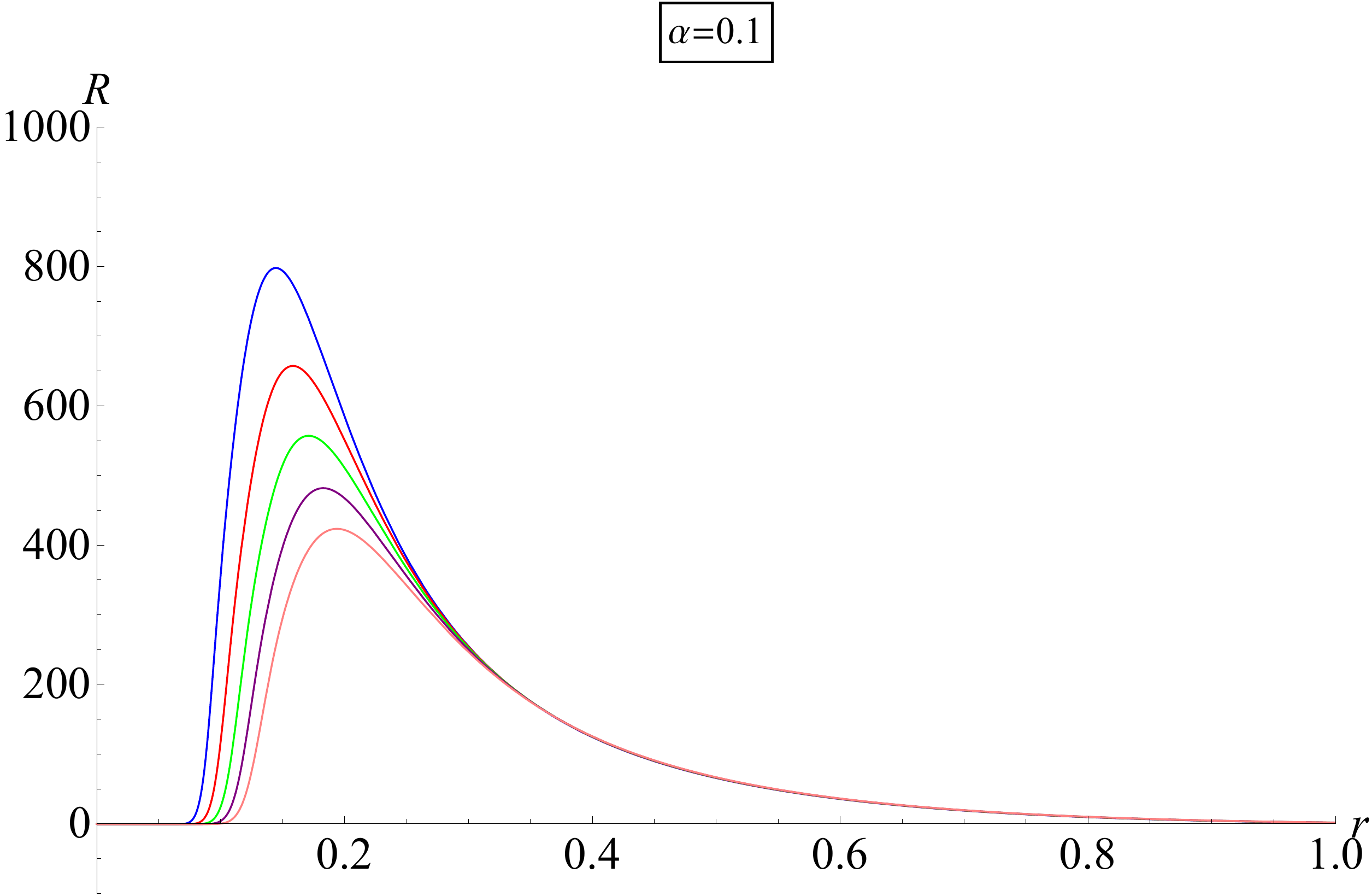}
\hspace*{0.05\textwidth}
\includegraphics[width=0.45 \textwidth]{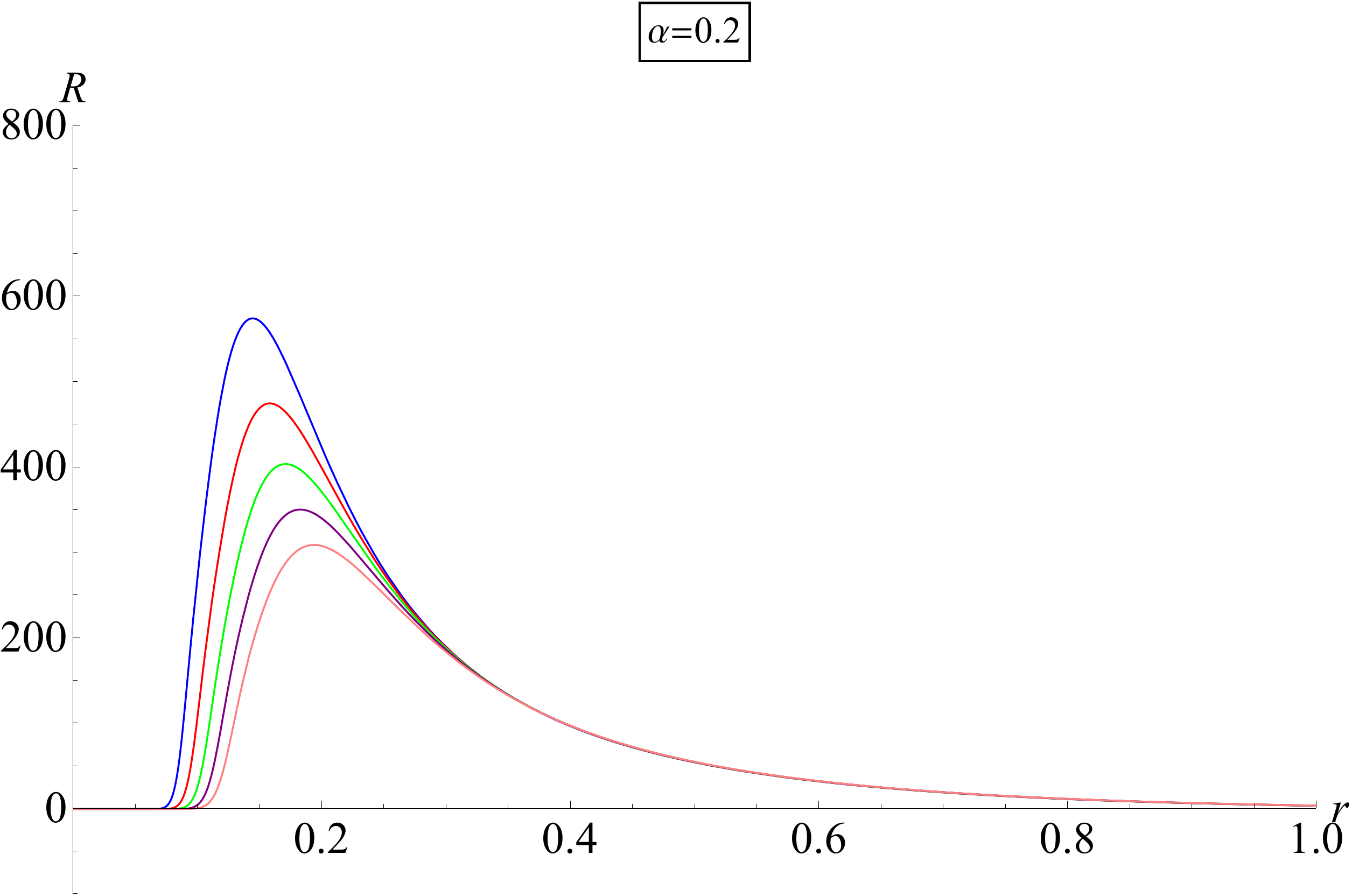}\\
\includegraphics[width=0.45 \textwidth]{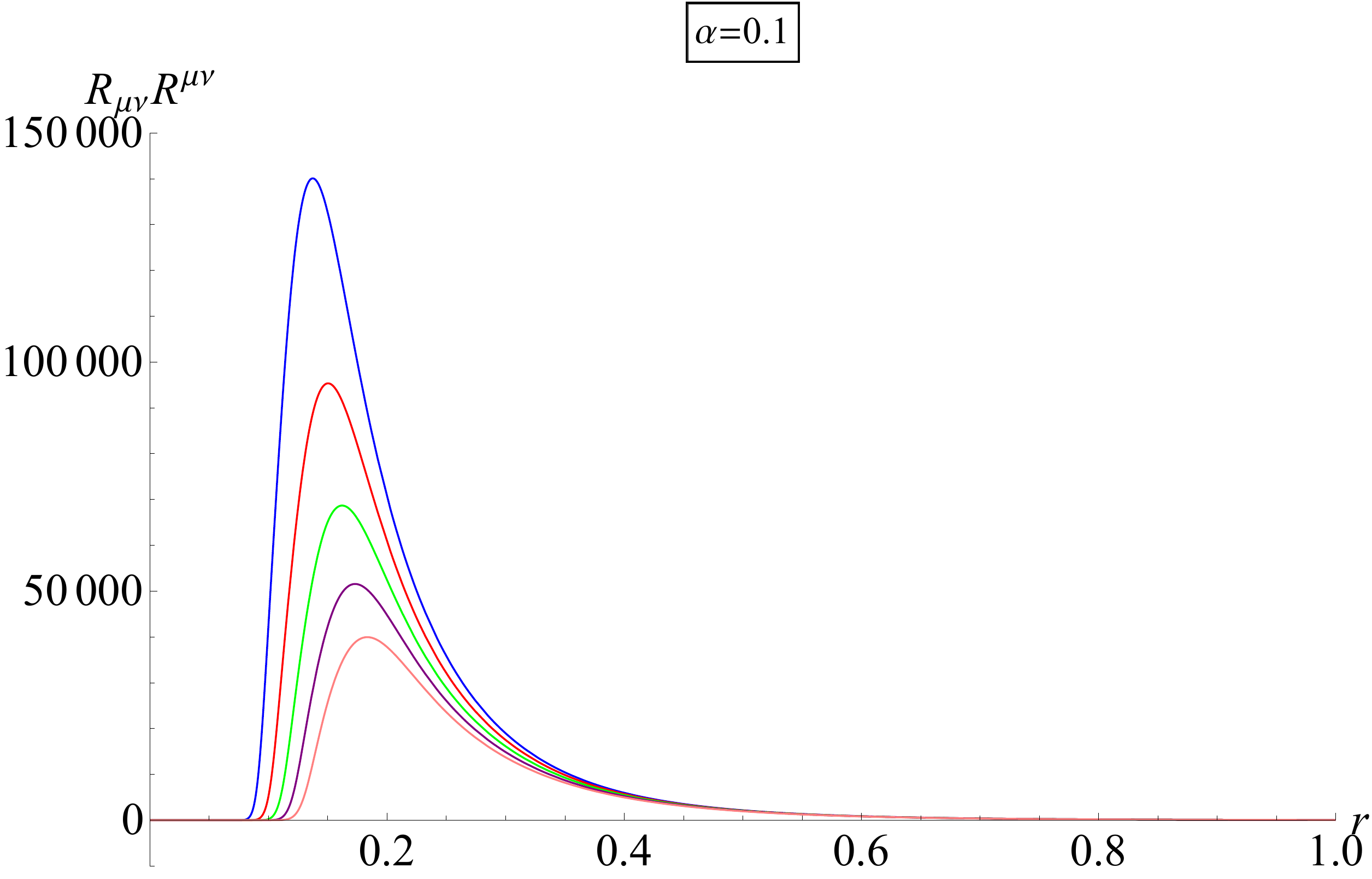}
\hspace*{0.05\textwidth}
\includegraphics[width=0.45 \textwidth]{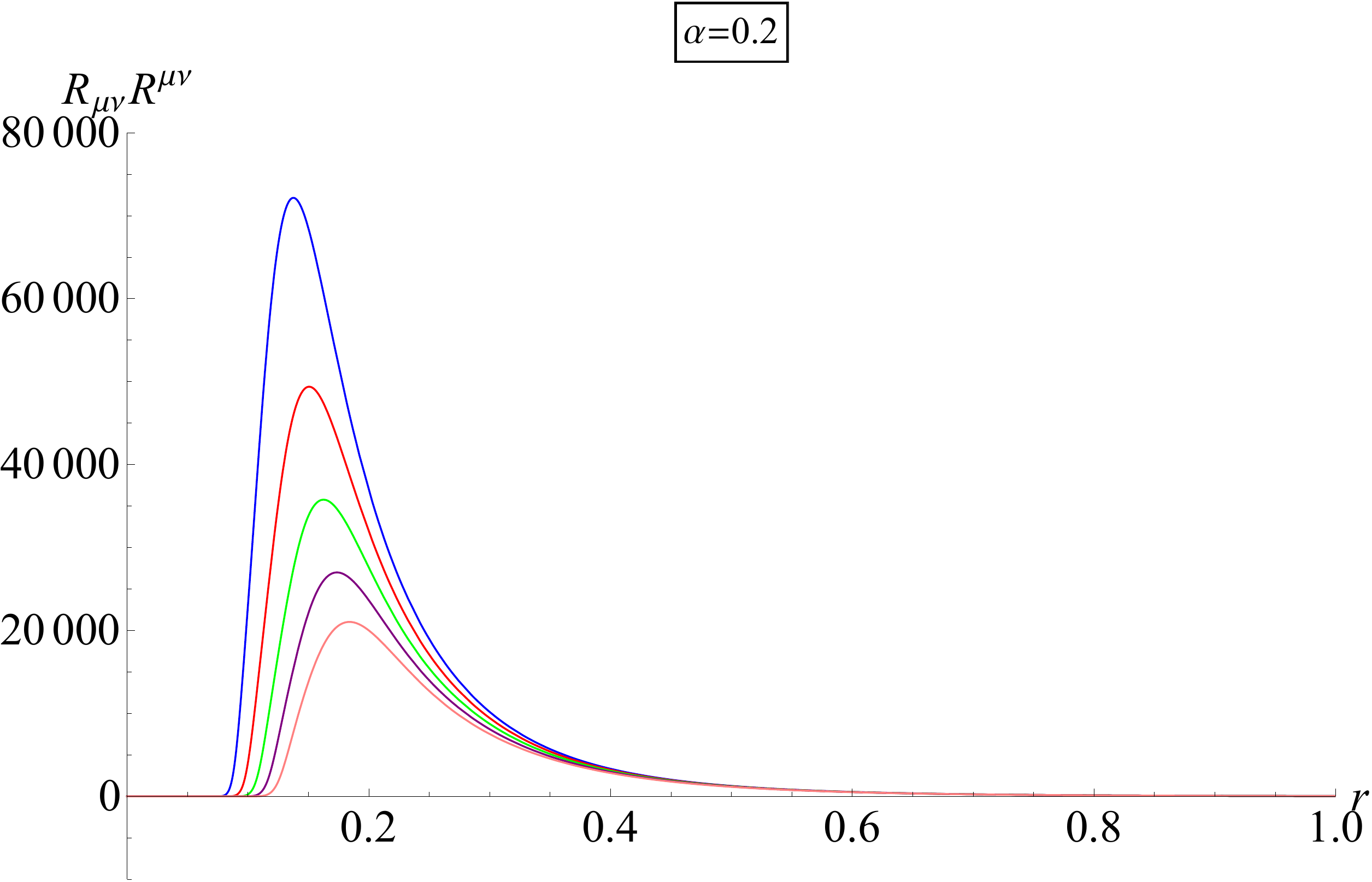}\\
\includegraphics[width=0.45 \textwidth]{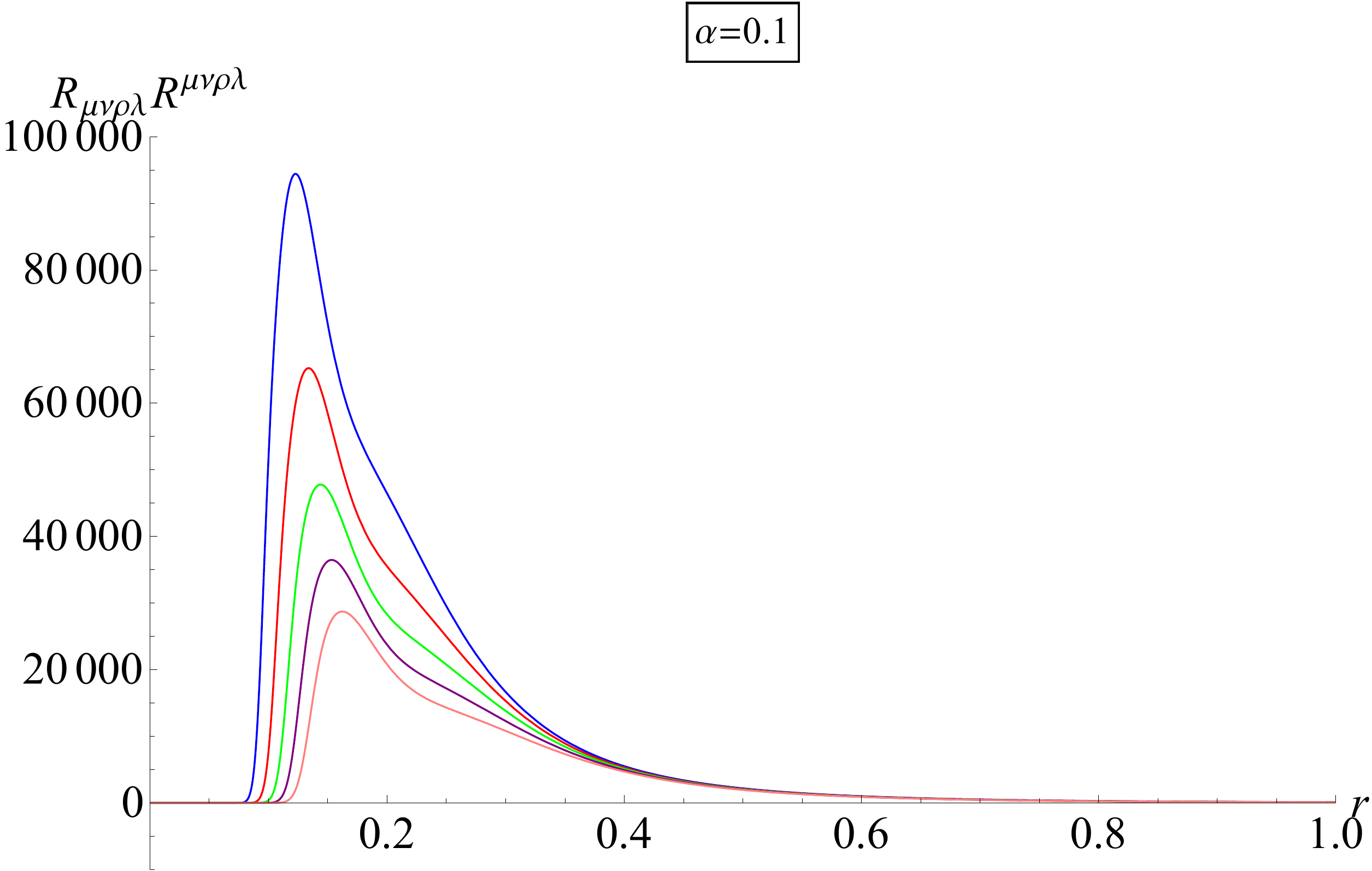}
\hspace*{0.05\textwidth}
\includegraphics[width=0.45 \textwidth]{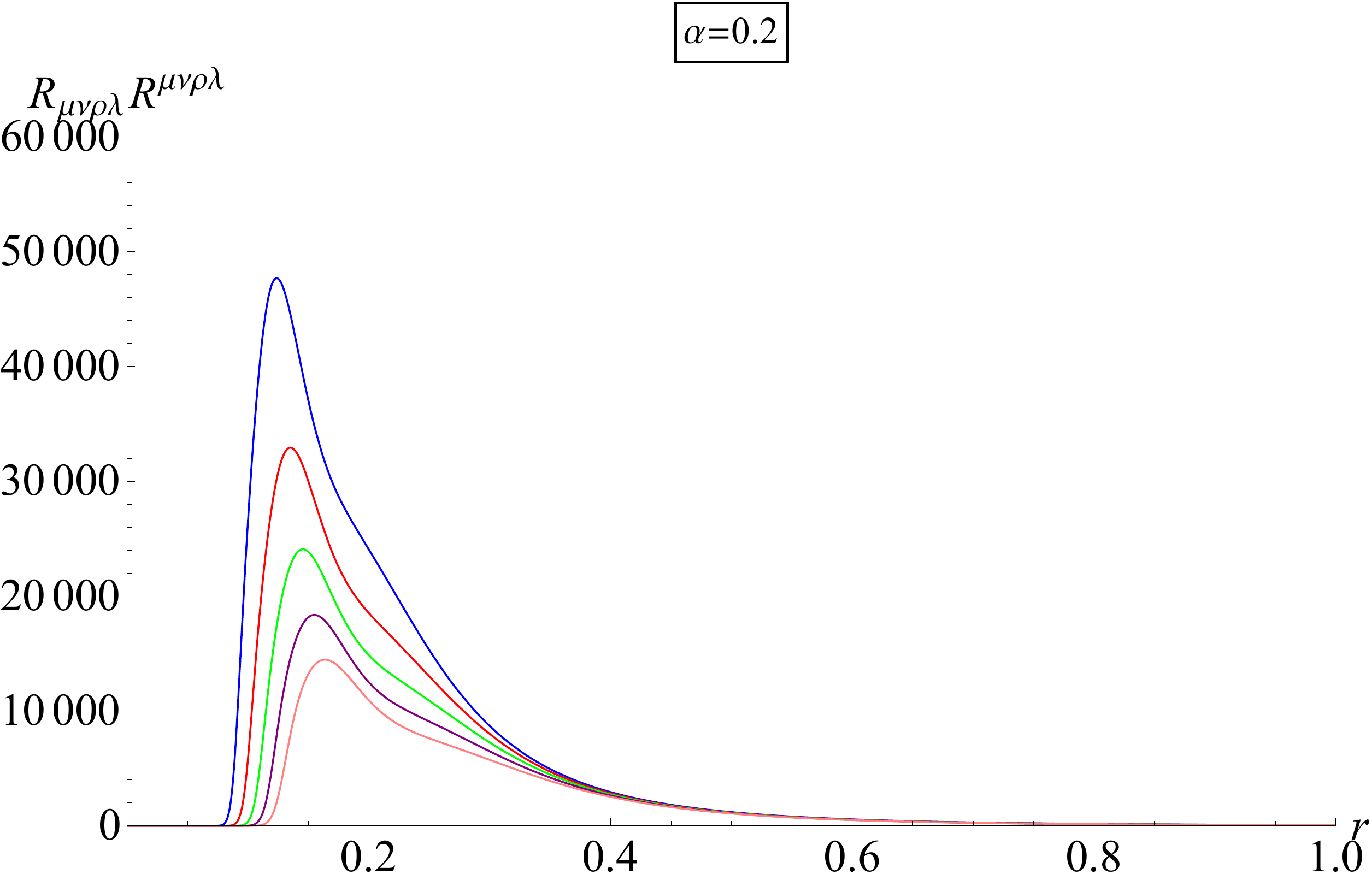}
\end{tabular}
  \caption{Plots of the curvature scalars $R$, $R_{\mu\nu}R^{\mu\nu}$, and
$R_{\mu\nu\rho\lambda}R^{\mu\nu\rho\lambda}$ in terms of $r$, under the different values of $k$ and $\alpha$, at $\frac{q^2}{3k}=4$ and $l=5$. The blue, red, green, purple and pink curves correspond to $k=0.1$, $0.12$, $0.14$, $0.16$, $0.18$, respectively.}\label{cur-scalars}
\end{figure}

The behavior of the solutions depends on various parameters in the theory. In this section we describe the basic properties of the solutions, including the existence of curvature singularity.  Basically, for given coupling constants $\alpha$ and $k$ and the cosmological constant, there are two classes of the solutions. The first one is when the mass is in the region 
, $m > \frac{q^{2}}{3k}$. In this class, the behavior of the curvature scalar $R$ near $r=0$ is given by,
\begin{equation}\label{sca-cur}
R=-\frac{5}{\alpha}+\frac{3}{r^2}\sqrt{\frac{2}{\alpha}\left(m-\frac{q^2}{3k}\right)}+\mathcal{O}\left(r^2\right),
\end{equation}
which indicates the existence of a curvature singularity at the center of the black hole. It describes the usual black hole solutions with curvature singularity inside the event horizon.
The other one is when the  mass saturates the lower bound $m=\frac{q^{2}}{3k}$, in which the curvature singularity disappears  and thus the black hole becomes regular black hole. In this case, the metric function is given by
\begin{eqnarray}\label{reg-bh}
f(r)=1+\frac{r^2}{4\alpha}\left(1-\sqrt{1-\frac{8\alpha}{l^2}+\frac{8\alpha q^{2}}{3kr^4}e^{-\frac{k}{r^2}}}\right).
\end{eqnarray}
In order to see the regularity of the solution, let us consider the behavior of the black hole at the short distance. In the limit $r\rightarrow0$, the function $f(r)$ behaves as
\begin{equation}
  f(r)\rightarrow 1+\frac{r^2}{l_{\rm eff}^{2}}.
\end{equation}
This means that the inside of the black hole including the curvature singularity at the origin is replaced by the core of the AdS geometry, and thus the black hole is regular. 
Indeed, this behavior is confirmed as various curvature invariants are smooth as $r\rightarrow0$:
\begin{eqnarray}
R\rightarrow -\frac{20}{l_{\rm eff}^{2}},\qquad
R_{\mu\nu}R^{\mu\nu}\rightarrow \frac{80}{l_{\rm eff}^{4}},\qquad
R_{\mu\nu\rho\lambda}R^{\mu\nu\rho\lambda}\rightarrow\frac{40}{l_{\rm eff}^{4}}.
\end{eqnarray}
 In addition, we can see that from Fig. \ref{cur-scalars} these curvature scalars are well-behaved everywhere and the black hole is indeed regular.

An essential property of the regular black hole solutions is that their interior near $r=0$ looks like the vacuum solution. As just indicated, the core region of our regular black hole solution is the same as the one of AdS geometry, just like the Ayon-Beato-Garcia regular black hole \cite{Ayon-Beato98,Schee2015}. Whereas, the standard Bardeen solution \cite{Bardeen,Ayon-Beato2000} and the regular black hole solutions obtained by Dymnikova \cite{Dymnikova2004,Dymnikova1992,Dymnikova2003,Dymnikova2019} have the dS center.

Before we proceed, let us pause to clarify the differences between our charged AdS black hole solutions and those obtained in various gravity theories in the four- and higher-dimensional spacetime. Many charged AdS black hole solutions have been obtained in Einstein/modified gravity with/without the higher order derivative corrections of the $U(1)$ gauge field. For example, the charged AdS black hole solutions were obtained in Einstein gravity with power-law Maxwell field \cite{Gonzalez2009,Zangeneh2015,Dehghani2017,Rincon2107}, in Gauss-Bonnet gravity's rainbow \cite{Hendi2015,Hendi2016}, in Gauss-Bonnet gravity with power-law Maxwell field \cite{Hendi2009,Hendi2018b} or with quadratic nonlinear electrodynamics \cite{Panahiyan2016}, in third order Lovelock gravity with Born-Infeld type nonlinear electrodynamics \cite{Dehghani2015}, in third order quasitopological gravity with power-law Maxwell source \cite{Ghanaatian2019}, in Lovelock gravity coupled to Born-Infeld nonlinear electrodynamics \cite{Farhangkhah2018}, in Einsteinian cubic gravity \cite{Bueno2016}, and in the $f(R)$ gravity \cite{Olmo2011,ASheykhi2012}. One may expect that the higher order curvature corrections of quantum gravity as well as the higher order derivative corrections of the $U(1)$ gauge field can smooth the spacetime geometry. However, for all these black hole solutions, there is always a curvature singularity at $r=0$. The corrections of quantum gravitation as well as the nonlinear electrodynamics only lead to the modification of the usual black holes or decreasing the strength of the curvature singularity. In our black hole solutions, there exists a lower bound for the black hole mass where the combination of the corrections for both gravity and the electrodynamics can smooth the spacetime geometry completely. As a result, the curvature singularity disappears and the black hole corresponding to this lower bound is regular. The lower bound of the mass of the regular black hole is determined in terms of its charge and the nonlinear coupling. This relation is analogous to the one in the regular black hole solutions obtained in Refs. \cite{Toshmatov2018,Wang-Fan2016,Rodrigues2017,Toshmatov2017,Toshmatov2019}. One of the main virtues of our black hole solutions, in contrast to the above solutions, is that they contain not only regular black hole but also extremal black hole as a lower bound of mass, which gives a variety of scenarios for the end points of Hawking radiation.

Furthermore, for the regular black hole solutions mentioned in the introduction, the nonlinear coupling constant should be fine-tuned and related to the charge and mass of the regular black holes. In contrast to these regular black hole solutions, the nonlinear coupling constant in our theory is given as a free parameter and the usual black hole solutions with curvature singularity exist. The regular black hole emerges as a special limit of these solutions with the mass given in terms of the charge and the nonlinear coupling constant.

\begin{figure}[t]
 \centering
\begin{tabular}{cc}
\includegraphics[width=0.3 \textwidth]{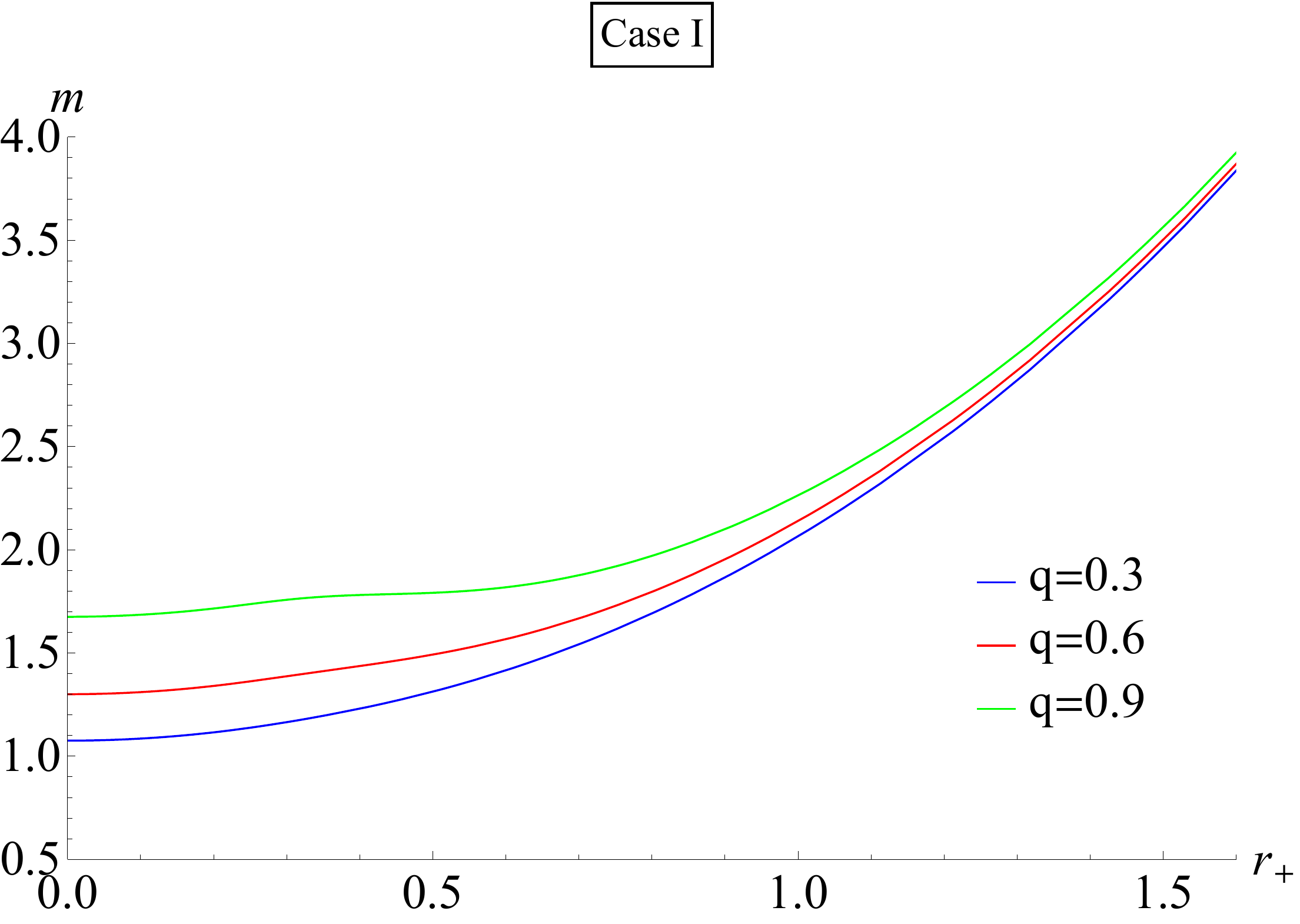}
\hspace*{0.01\textwidth}
\includegraphics[width=0.3 \textwidth]{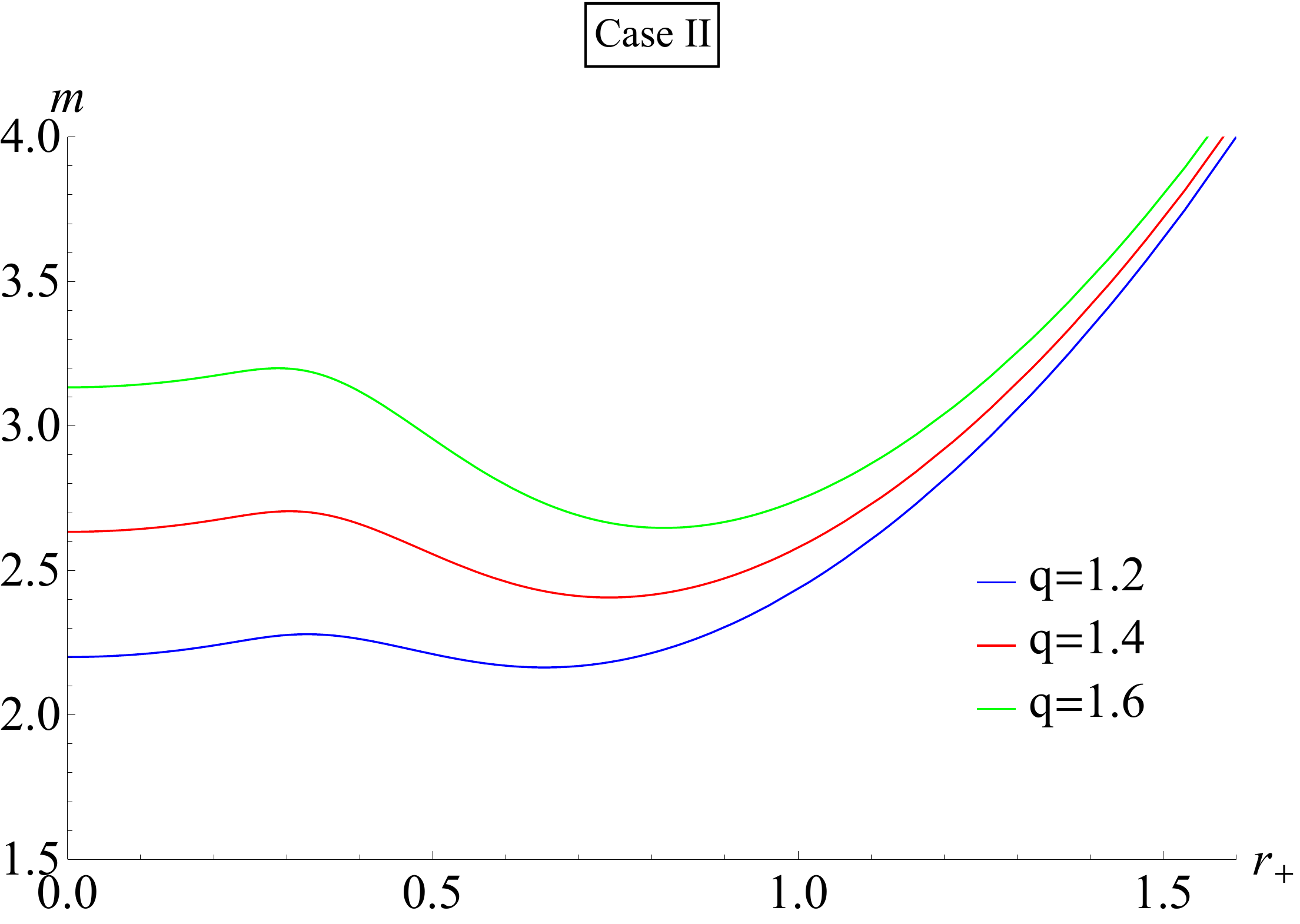}
\hspace*{0.01\textwidth}
\includegraphics[width=0.3 \textwidth]{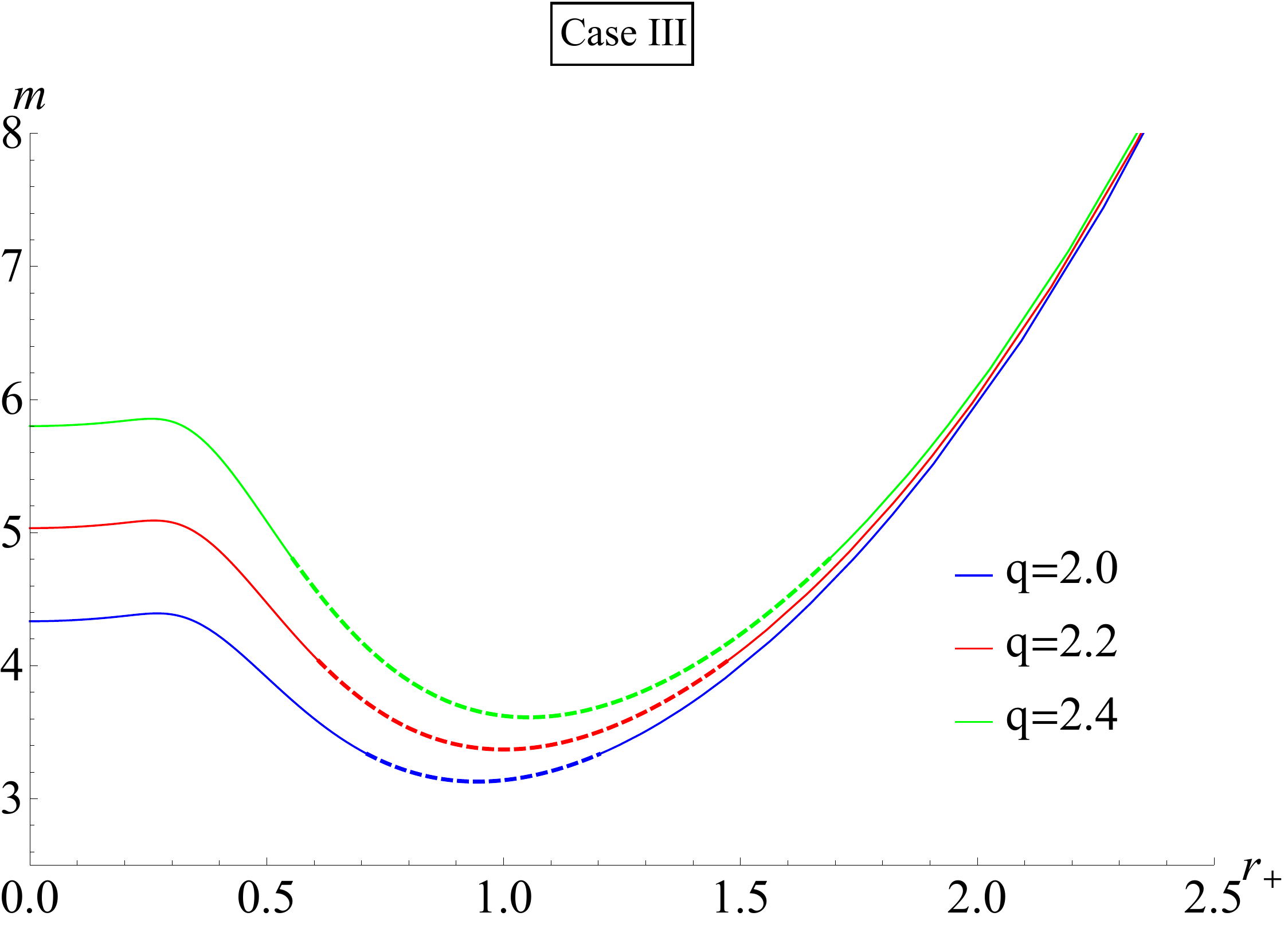}
\end{tabular}
  \caption{The mass function given at Eq. (\ref{massfunc}) is plotted in terms of the horizon radius, at $\alpha=0.5$, $l=5$ and $k=0.4$.}\label{mass-rad-A}
\end{figure}

From the equation of the horizon, $f(r_+)=0$, where $r_+$ denotes the horizon radius of the black hole, one can obtain the relation between the reduced black hole mass $m$ and the horizon radius $r_+$ as
\begin{equation}
  m=r^2_++2\alpha+\frac{r^4_+}{l^2}-\frac{q^2}{3k}\Big(e^{-\frac{k}{r^2_+}}-1\Big).\label{massfunc}
\end{equation}
One may note that the mass approaches $2\alpha+\frac{q^{2}}{3k}$ as $r_+\rightarrow0$ and grows indefinitely  as $r_+\rightarrow\infty$. As shown in Fig. \ref{mass-rad-A},  the behavior of the mass given at (\ref{massfunc}) as a function of the horizon radius crucially depends  on the parameters $k$, $\alpha$, $q$ and $l$. 
The first case is when the mass function $m(r_+)$ is a monotonically increasing function of $r_+$.  It is realized, for example, if  the charge $q$ is small enough  for given coupling constants $k$ and $\alpha$.\footnote{ Alternatively, it can be realized if the nonlinear gauge coupling $k$ is large enough for fixed $\alpha$ and the charge $q$.} More specifically, it corresponds to the case that the  charge $q$ is smaller than a critical value $q_c$ for given $l$ and $k$, where $q_c$ is determined in terms of $l$ and $k$ by the following relation\footnote{This relation is obtained from the condition that the maximum and minimum of $m(r_+)$ coincide together and become an inflection point.}
\begin{eqnarray}
q_{c}= r_{0}^{2}\sqrt{3\left(\frac{2r^2_0}{l^2}+1\right)e^{\frac{k}{r^2_0}}}, \qquad r^2_0\equiv\frac{1}{6}\left[k-l^2+\sqrt{(k-l^2)^2+6kl^2}\right].
\end{eqnarray}
 In this case, there is no local minimum of $m(r_+)$ and thus there is no extremal black hole solution.  Furthermore since the mass of the black hole is lower bounded as $2\alpha+\frac{q^{2}}{3k}$, which is larger than the mass of the regular black hole,  there is no  regular black hole solution either.  This would mean that the black hole can not reach the extremal black hole nor regular black hole through the Hawking radiation, in which the black hole can only lose the charge. This case is shown in the left panel of Fig. \ref{mass-rad-A} and will be denoted as the case I.

\begin{figure}[t]
 \centering
\begin{tabular}{cc}
\includegraphics[width=0.45 \textwidth]{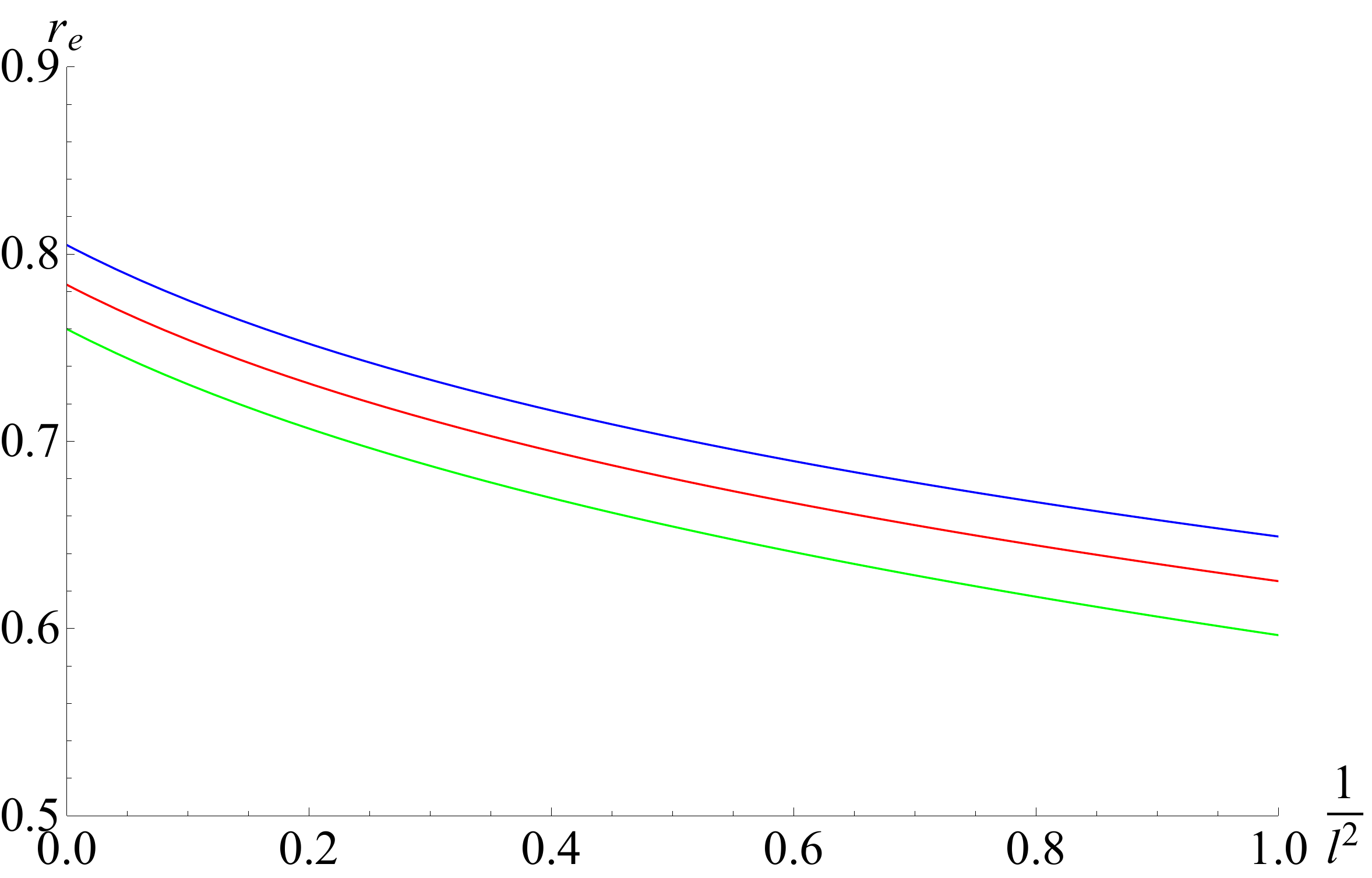}
\hspace*{0.05\textwidth}
\includegraphics[width=0.45 \textwidth]{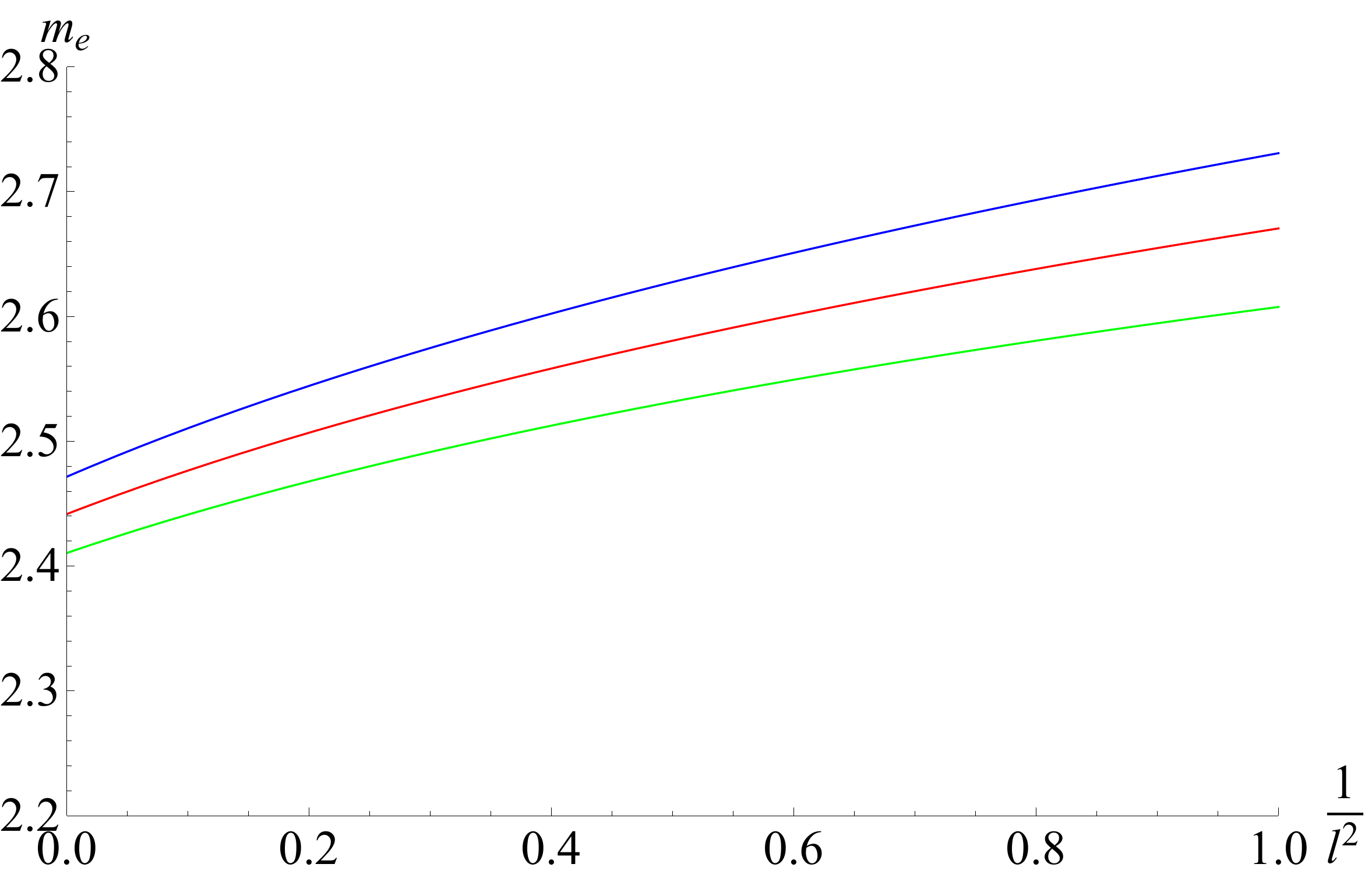}
\end{tabular}
  \caption{The extremal radius and the extremal mass are plotted in terms of the inverse squared curvature radius $\frac{1}{l^2}$, under the different values of $k$, at $\alpha=0.5$ and $q=\sqrt{2}$. The blue, red, and green curves correspond to $k=0.3$, $0.35$, $0.4$, respectively.}\label{extre-rad-mass}
\end{figure}

As we increase the charge $q$, the mass function develops the local minimum, which corresponds to the extremal black hole. The  horizon radius $r_e$ of the extremal black hole is a large positive root of the equation $F_{1}(r_{e})=0$, where
\begin{equation}
F_1(r_{+})\equiv  \frac{2r^6_+}{l^2}+r^4_+-\frac{q^2}{3}e^{-\frac{k}{r^2_+}},\label{ext-eq}
\end{equation}
and is graphically solved in the left panel of Fig. \ref{extre-rad-mass}.
This figure suggests that increasing the nonlinear parameter $k$ makes both extremal radius and extremal mass decreasing. On the other hand, increasing  the cosmological constant leads to  the decrease of the extremal radius and the increase of the mass. 
In this situation, we have two different cases. For intermediate region of the charge $q$, the reduced mass of the extremal black hole is larger than the one of regular black hole $m=\frac{q^{2}}{3k}$. In this case the regular black hole solution may never be reached by the Hawking radiation and the extremal black hole would be the end point of the radiation. It is depicted in the middle panel of  Fig. \ref{mass-rad-A} and will be called the case II. 

On the other hand, if  the charge $q$ is  large enough  for given coupling constants $k$ and the cosmological constant, the reduced mass of extremal black hole becomes smaller than the reduced mass of the regular black hole.  It is depicted in the right panel of  Fig. \ref{mass-rad-A} and will be denoted as the case III. Since the regular black hole mass is larger than the global minimum of the curve $m(r_+)$, there is no  black hole solutions with the mass in the dashed part. One may note that the horizon radius of the regular black hole is the root of the  equation,
\begin{eqnarray}
r^2_++2\alpha+\frac{r^4_+}{l^2}-\frac{q^2}{3k}e^{-\frac{k}{r^2_+}}=0.
\end{eqnarray}
In this case we have interesting possibility for the fate of the black hole.  The regular black hole may be reached first by the  Hawking radiation and then, since the regular black hole has non-zero Hawking temperature, it would still radiate. Depending on the amount of the loss of the charge $q$ through the radiation, there is a chance that it ends up  to be regular as well as extremal.

\section{\label{therm} Thermodynamics and Phase Transitions}

In the extended phase space, the cosmological constant plays the role of the pressure as
\begin{equation}
P=-\frac{\Lambda}{8\pi}=\frac{3}{4\pi l^2},
\end{equation}
in which the mass of the black hole is naturally regarded as the enthalpy. 
 In this description, the first law of the black hole thermodynamics is given by
\begin{eqnarray}\label{1stlaw}
dM=TdS+\Phi dQ+VdP, 
\end{eqnarray}
where the entropy $S$, the total charge $Q$ and the pressure $P$ are a complete set of the thermodynamic variables for the mass or enthalpy function $M(S,Q, P)$, while the temperature $T$, the chemical potential $\Phi$ and the thermodynamic volume $V$ are conjugate variables to $S$, $Q$ and $P$, respectively, and are defined as
\begin{equation}
T=\left(\frac{\partial M}{\partial S}\right)_{Q,P},\ \ \ \
\Phi=\left(\frac{\partial M}{\partial Q}\right)_{S,P},\ \ \ \ 
V=\left(\frac{\partial M}{\partial P}\right)_{S,Q}.
\end{equation}

\begin{figure}[t]
 \centering
\begin{tabular}{cc}
\includegraphics[width=0.3 \textwidth]{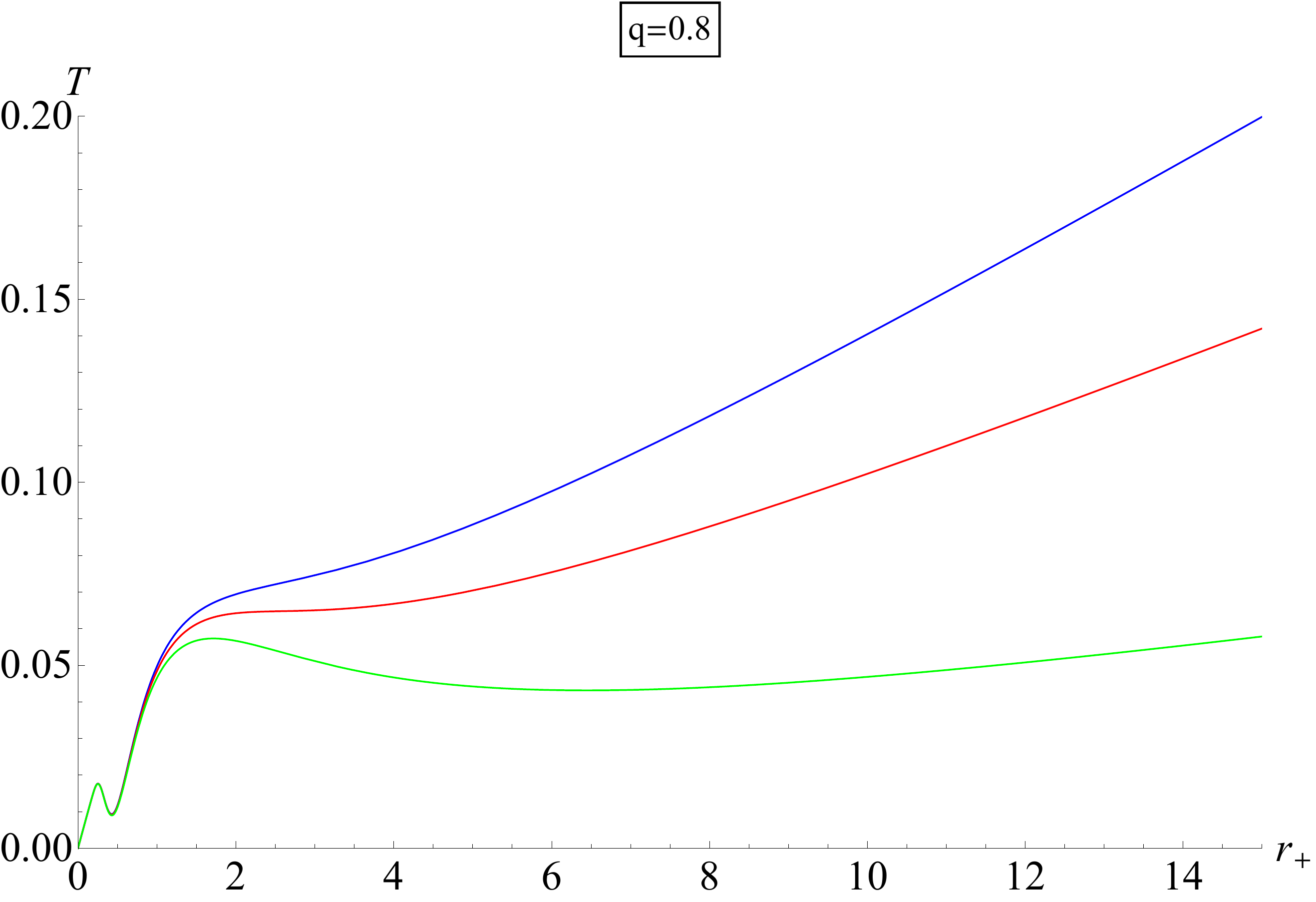}
\hspace*{0.01\textwidth}
\includegraphics[width=0.3 \textwidth]{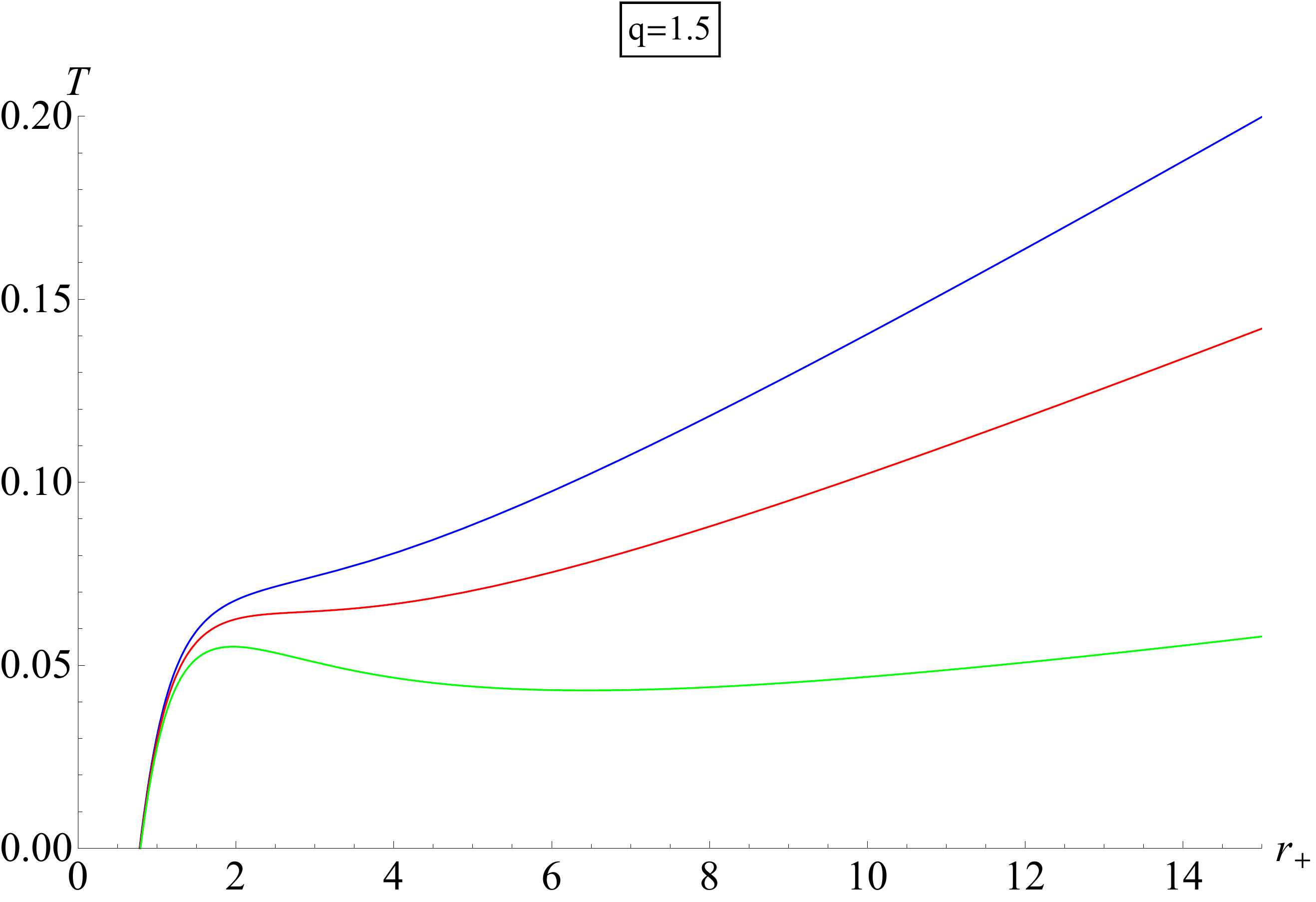}
\hspace*{0.01\textwidth}
\includegraphics[width=0.3 \textwidth]{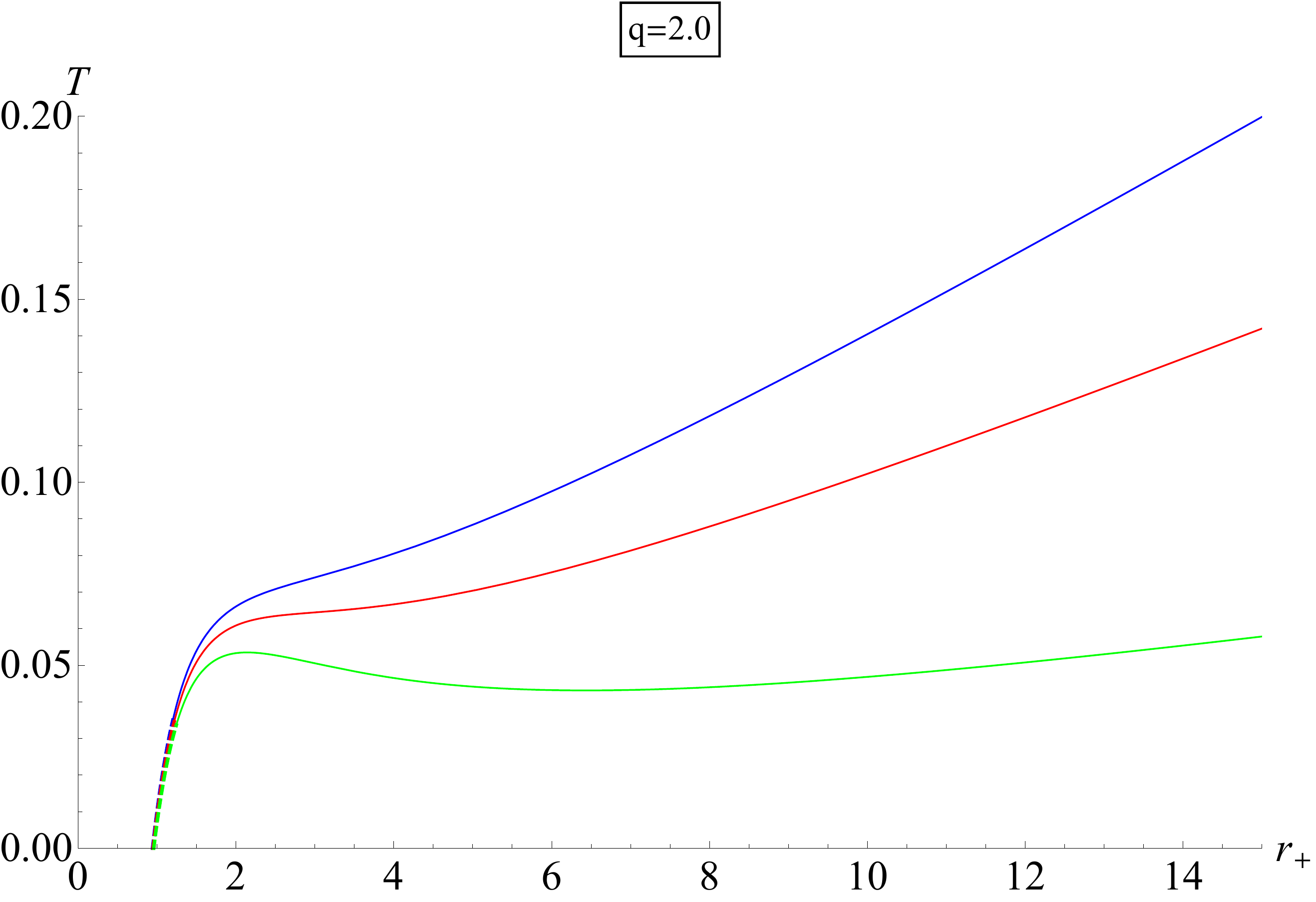}
\end{tabular}
  \caption{The isobaric curves in the $T-r_+$ diagram are plotted for various values of the charge $q$ and the pressure $P$, at $\alpha=0.5$ and $k=0.4$. The blue, red, and green curves correspond to $P=\frac{3}{100\pi}$, $\frac{1}{48\pi}$, $\frac{3}{400\pi}$, respectively. There is no black hole solution in the dashed part.}\label{bh-temp}
\end{figure}

The Bekenstein-Hawking temperature can be obtained from the surface gravity at the event horizon as \cite{Gibbons1977a,Gibbons1977b}
\begin{eqnarray}
T=\frac{f'(r_+)}{4\pi}=\frac{r^4_+\left(8\pi Pr^2_++3\right)-q^2e^{-\frac{k}{r^2_+}}}{6\pi r^3_+\left(r^2_++4\alpha\right)}.\label{bhtemp}
\end{eqnarray}
The behavior of the black hole temperature as a function of the radius of event horizon is plotted for the cases I (left panel), II (middle panel) and III (right panel) in Fig. \ref{bh-temp}, which correspond to small, intermediate and large charge $q$, respectively.
Those isobaric  curves have local extrema, which would signal phase transitions.  The radii $r_{\rm ex}$ corresponding to the local extremum temperatures are solutions of the equation $F_{2}(r_+)=0$ where
\begin{equation}
F_2(r_+)\equiv  r^6_+\left\{12\alpha+r^2_+\left[8\pi P(r^2_++12\alpha)-3\right]\right\}+q^2\left[5r^4_++12r^2_+\alpha-2k(r^2_++4\alpha)\right]e^{-\frac{k}{r^2_+}}.\label{max-min-equ}
\end{equation}
Some of those extrema may disappear if the pressure is larger than a certain critical value $P_{c}$, where
 the maximal and minimal temperature points coincide together, i.e.,
\begin{equation}
\left(\frac{\partial T}{\partial
r_+}\right)_{P_{c}}=\left(\frac{\partial^2T}{\partial r^2_+}\right)_{P_{c}}=0.
\end{equation}
We give the numerical
results of the equation $F_{2}(r_{\rm ex})=0$ in Table \ref{min-max-sols}.
\begin{table}[!htp]
\centering
\begin{tabular}{|c|c|c|c|c|c|c|c|c|c|}
  \hline
  \multicolumn{5}{|c|}{$\alpha=0.5$} & \multicolumn{5}{|c|}{$\alpha=1$} \\
  \hline
  $P$ & $r_{\textrm{max}}$ & $T_{\textrm{max}}$ & $r_{\textrm{min}}$ & $T_{\textrm{min}}$ & $P$ & $r_{\textrm{max}}$ & $T_{\textrm{max}}$ & $r_{\textrm{min}}$ & $T_{\textrm{min}}$\\
  \hline
  0.001 & 1.84532 & 0.05307 & 10.5445 & 0.02864 & 0.0005 & 2.28472 & 0.0394 & 14.9138 & 0.02025 \\ \hline
  0.002 & 1.90893 & 0.05466 & 7.16091 & 0.03976 & 0.001 & 2.36655 & 0.0403 & 10.133 & 0.02812 \\ \hline
  0.003 & 1.98842 & 0.05636 & 5.57573 & 0.04776 & 0.0015 & 2.46662 & 0.04124 & 7.89979 & 0.03378 \\ \hline
  0.004 & 2.0948 & 0.0582 & 4.55215 & 0.05398 & 0.002 & 2.59544 & 0.04228 & 6.47005 & 0.0382 \\ \hline
  \end{tabular}
\caption{The numerical results for the local maximum temperature $T_{\textrm{max}}$ and the local minimum temperature $T_{\textrm{min}}$ corresponding to the radius $r_{\textrm{max}}$ and the radius $r_{\textrm{min}}$ are given in terms of the pressure under various values of Gauss-Bonnet coupling constant $\alpha$, at $k
=0.35$ and $q=\sqrt{2}$.}\label{min-max-sols}
\end{table}

\begin{figure}[t]
 \centering
\begin{tabular}{cc}
\includegraphics[width=0.45 \textwidth]{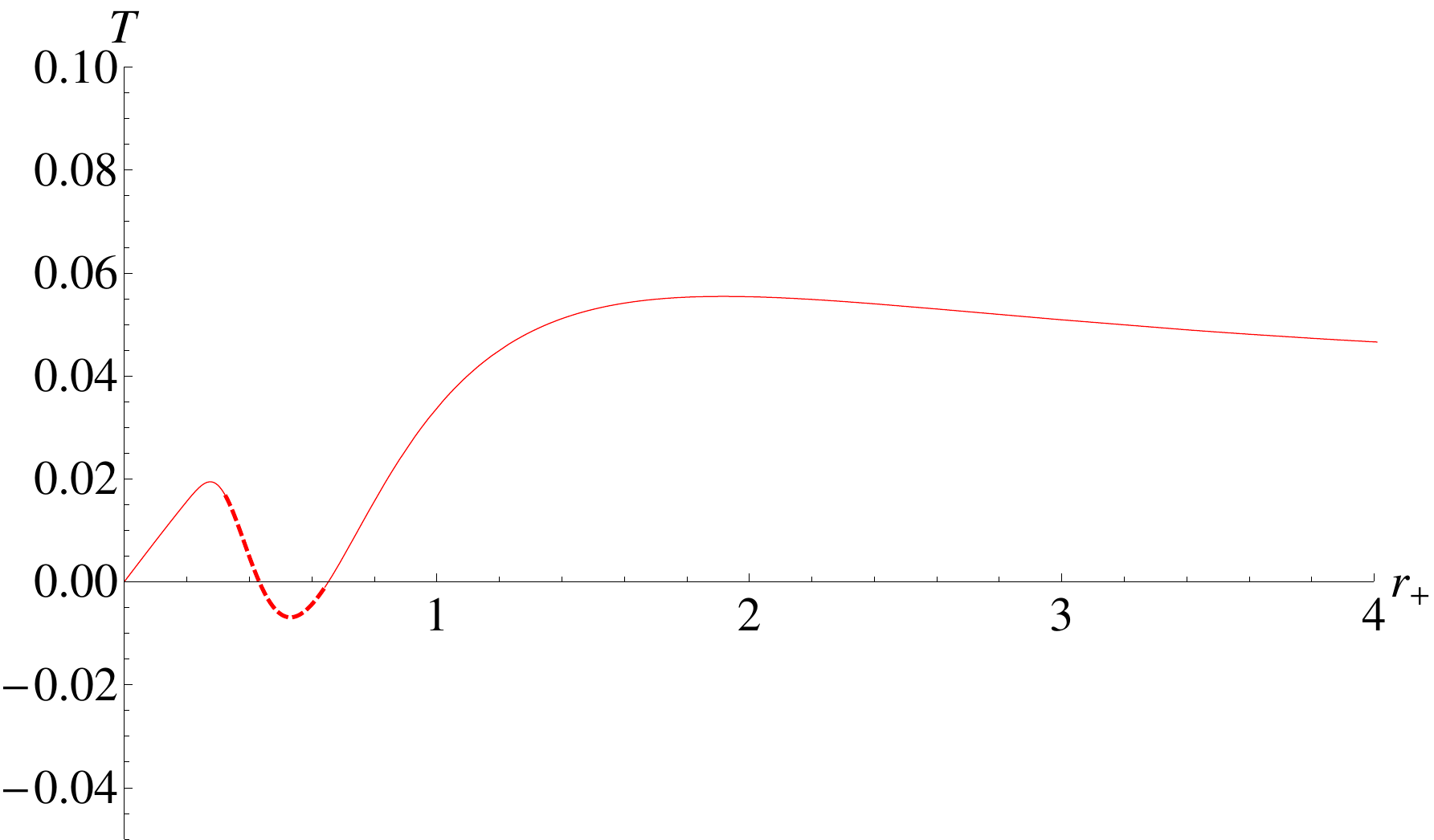}
\end{tabular}
  \caption{The isobaric curve for the case II is plotted at $\alpha=0.5$, $k=0.55$, $P=\frac{3}{100\pi}$ and $q=\sqrt{2}$. There is no the black hole solution  in the dashed part.}\label{spe-casI}
\end{figure}

It is worthwhile to note that, in the case II, the curve of the black hole temperature  has, for some range of parameters,  two allowed regions separated by a forbidden region, in which there is no corresponding black hole solution, as described in Fig. \ref{spe-casI}. 
It means that the black hole solution  remains in one of these allowed regions throughout the Hawking radiation, depending on the initial condition.  

The black hole entropy can be obtained from the Wald formula and, as shown in the appendix, is given by
\begin{eqnarray}
  S 
  &=&\frac{S_3r^3_+}{4}\left(1+\frac{12\alpha}{r^2_+}\right),\label{bhent}
\end{eqnarray}
where the second term is the additional contribution from Gauss-Bonnet term away from the usual area law for the Einstein gravity. Surely, this can also be confirmed by the integration of 
the first law as
\begin{eqnarray}
  S &=& \int\frac{1}{T}\frac{\partial M}{\partial r_+}dr_+=\frac{3S_3}{4}\int\left(r^2_++4\alpha\right)dr_+
  =\frac{S_3r^3_+}{4}\left(1+\frac{12\alpha}{r^2_+}\right).\label{bhent}
\end{eqnarray}

The chemical potential $\Phi$ is  obtained from the first law as
\begin{eqnarray}
\Phi&=&\left(\frac{\partial M}{\partial Q}\right)_{S,P}=\frac{q}{6}\left(\frac{2}{k}-\frac{2r^2_+-k}{kr^2_+}e^{-\frac{k}{r^2_+}}\right),\label{chem}
\end{eqnarray}
which is nothing but the electrostatic potential given in Eqs. (\ref{potential}).
The  thermodynamic volume $V$ can also be obtained from the first law as
\begin{eqnarray}
V&=&\left(\frac{\partial M}{\partial P}\right)_{S,Q}=V_4r^4_+,\label{thervol}
\end{eqnarray}
where $V_{4}=\frac{\pi^{2}}{2}$ is the volume of unit 4-ball in the flat space. 

Before we move to the investigation of phase transitions, let us pause to  take the  limit $l\rightarrow\infty$, where we obtain black hole solutions  without cosmological constant. In this case, the black hole mass given in Eqs. (25) and (34) becomes
\begin{eqnarray}
M=\frac{3S_3}{16\pi}\left[r^2_++2\alpha-\frac{q^2}{3k}\Big(e^{-\frac{k}{r^2_+}}-1\Big)\right],
\end{eqnarray}
 and the black hole temperature becomes
\begin{eqnarray}
T=\frac{3r^4_+-q^2e^{-\frac{k}{r^2_+}}}{6\pi r^3_+\left(r^2_++4\alpha\right)}.
\end{eqnarray}
The expressions of the black hole entropy $S$ and the chemical potential $\Phi$ are not affected by the cosmological constant and  remain the same. The first law of the black hole thermodynamics is  satisfied as
\begin{eqnarray}
dM=TdS+\Phi dQ.
\end{eqnarray}

The thermodynamic stability of the black hole is determined by its heat capacity $C_P$ which is given by
\begin{eqnarray}
C_P&=&T\left(\frac{\partial S}{\partial T}\right)_P=\frac{\partial M}{\partial r_+}\left(\frac{\partial T}{\partial r_+}\right)^{-1}
=\frac{9S_3}{4}\frac{r^3_+(r^2_++4\alpha)^2F_1(r_+)}{F_2(r_+)},
\end{eqnarray}
where $F_1(r_+)$ and $F_2(r_+)$ are given in Eqs. (\ref{ext-eq}) and (\ref{max-min-equ}), respectively.
\begin{figure}[t]
 \centering
\begin{tabular}{cc}
\includegraphics[width=0.45 \textwidth]{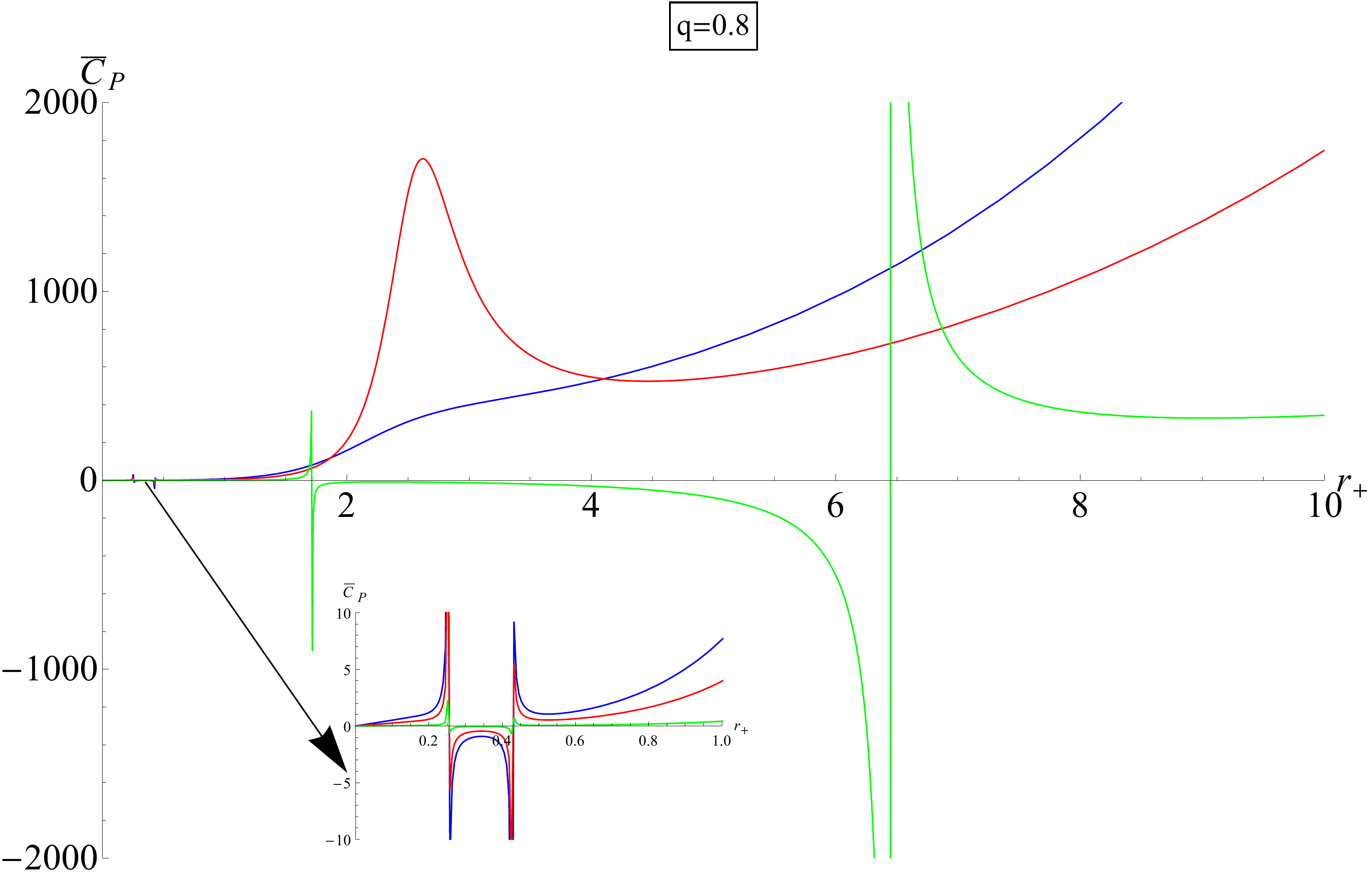}
\hspace*{0.05\textwidth}
\includegraphics[width=0.45 \textwidth]{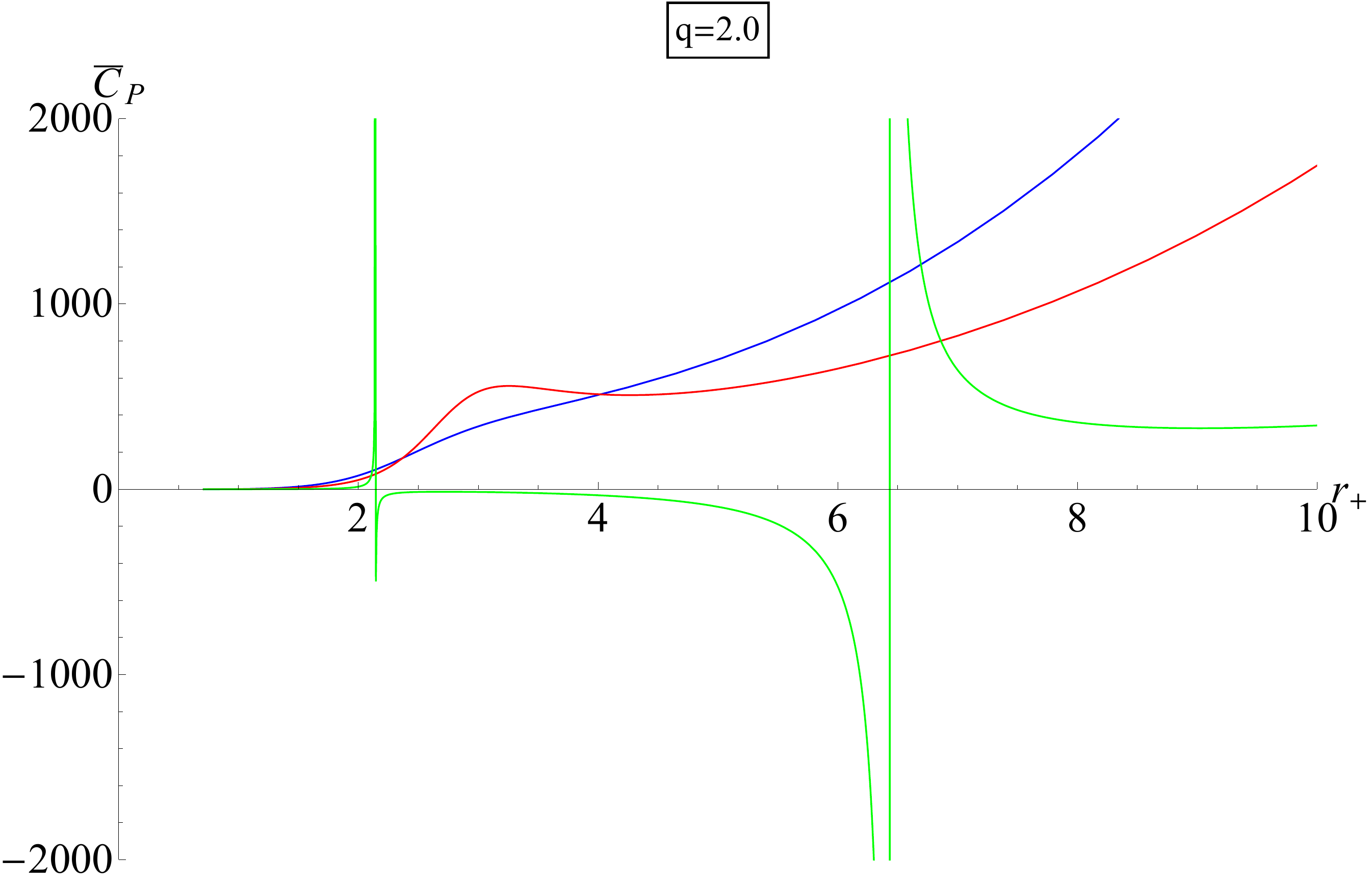}
\end{tabular}
  \caption{Plots of the reduced heat capacity $\bar{C}_P=\frac{4}{9S_3}C_P$ of the black hole, at $\alpha=0.5$ and $k=0.4$. The blue, red, and green curves correspond to $P=\frac{3}{100\pi}$, $\frac{1}{48\pi}$, $\frac{3}{400\pi}$, and are scaled by $\frac{1}{5}$, $\frac{1}{10}$, $\frac{1}{100}$, respectively.}\label{heat-cap}
\end{figure}
If the heat capacity $C_P$ is positive, the black hole is thermodynamically stable and if it is negative, the black hole is unstable. The black hole undergoes the second-order phase transition between these stable and unstable phases. This is due to the fact that the heat capacity $C_{P}$ suffers from discontinuity at the extremal points $r_{\rm ex}$ satisfying $F_{2}(r_{\rm ex})=0$. In Fig. \ref{heat-cap}, we plot the black hole heat capacity in terms of the radius of the event horizon under various values of the pressure and the charge $q$. 

The black holes for the case I with the proper charge always have the thermodynamically unstable region while  the black holes for the cases II and III do not have one for the pressure above the critical pressure. On the other hand, for the pressure below the critical pressure, the black holes in all three cases have a thermodynamically unstable phase between two thermodynamically stable phases. 

\begin{figure}[t]
 \centering
\begin{tabular}{cc}
\includegraphics[width=0.45 \textwidth]{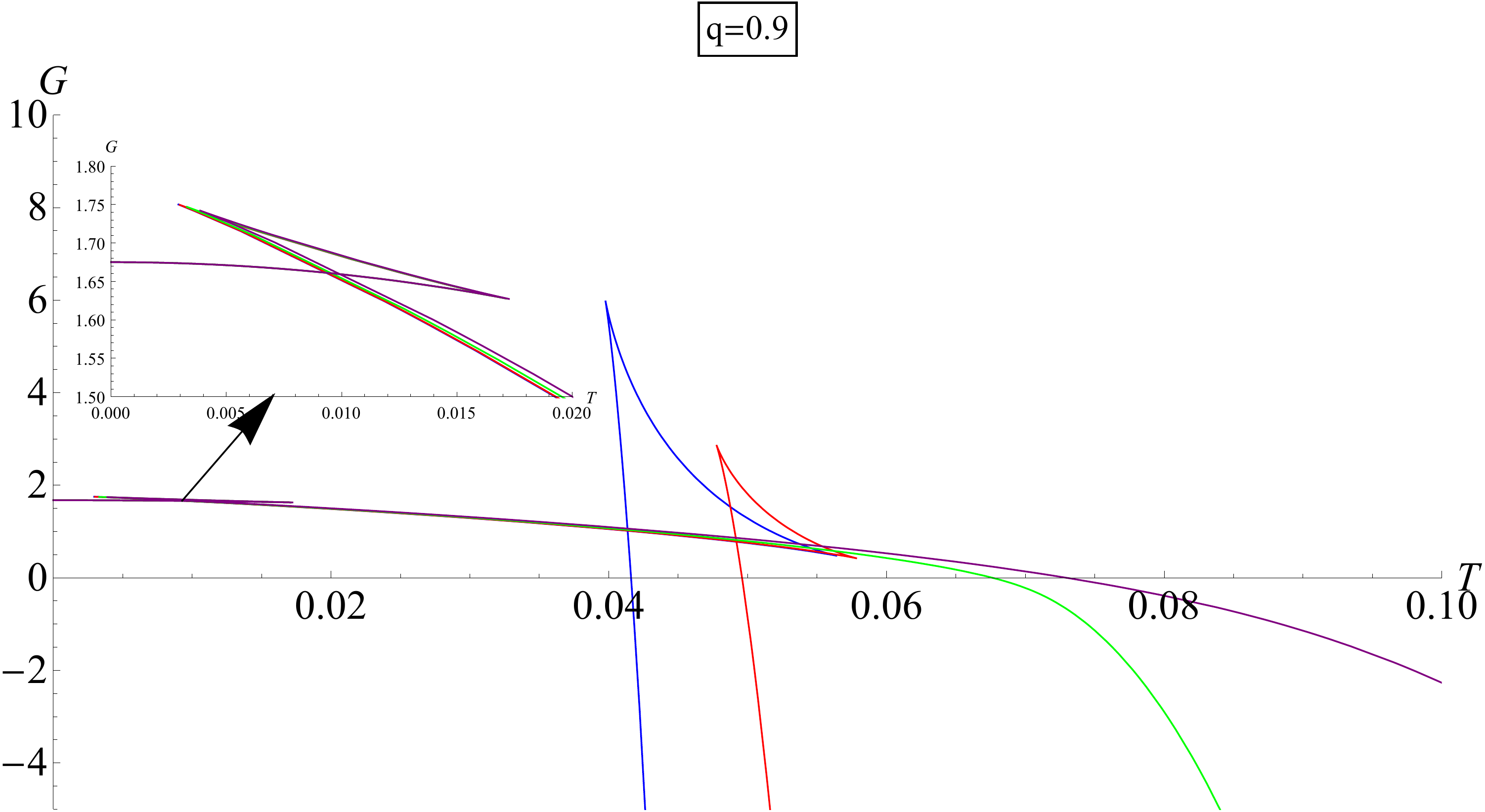}
\hspace*{0.05\textwidth}
\includegraphics[width=0.45 \textwidth]{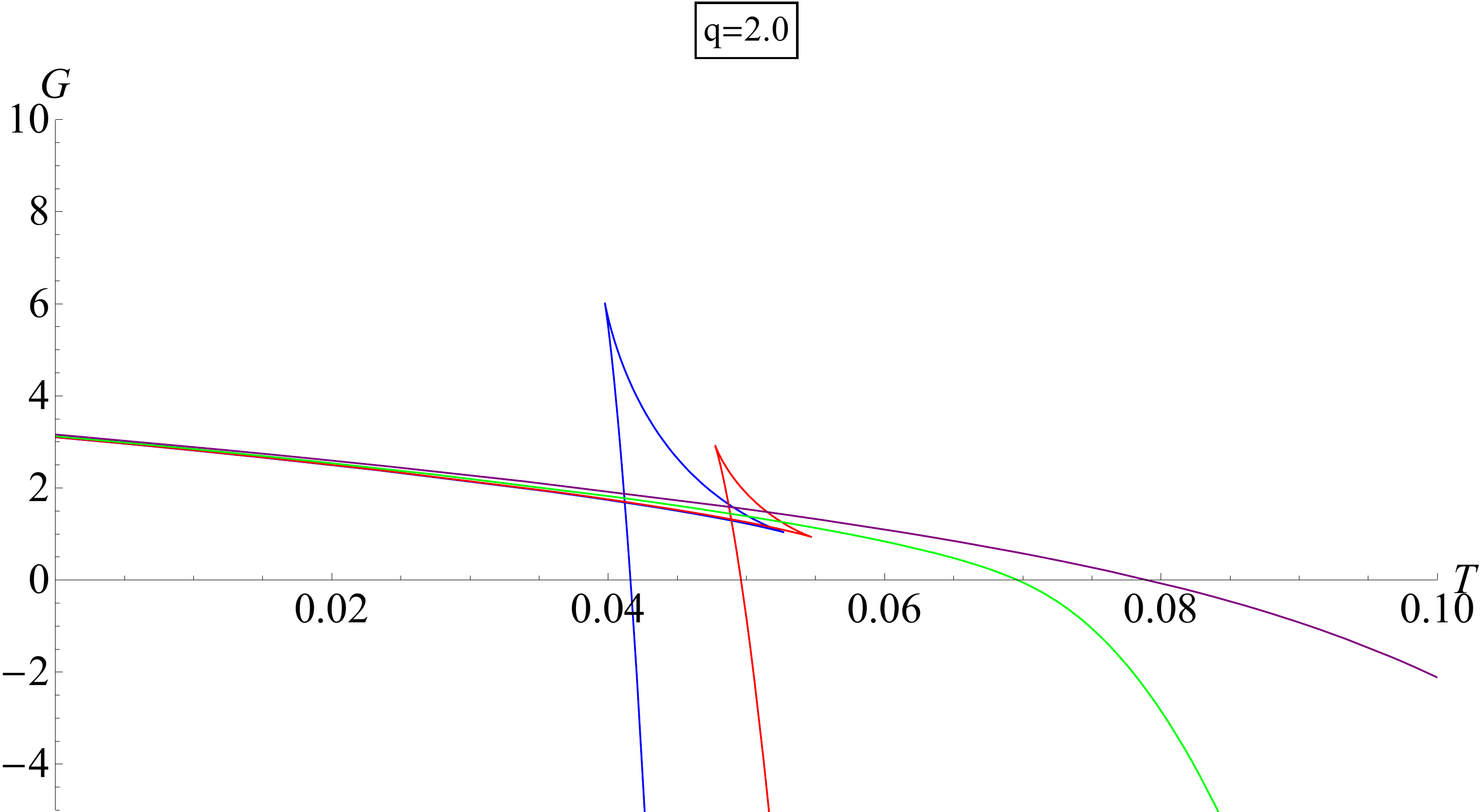}
\end{tabular}
  \caption{Plots of the Gibbs free energy, scaled by $16\pi/S_3$, in terms of the event horizon radius, at $\alpha=0.5$ and $k=0.4$. The blue, green, red, and purple curves correspond to $P=0.002$, $0.003$, $0.009$, $0.02$, respectively.}\label{Gibbs-energy}
\end{figure}

By analyzing the Gibbs free energy $G$, we can obtain another kind of phase transition.
The Gibbs free energy $G$ is defined by 
\begin{equation}
G=M-TS=\frac{S_3}{16\pi}\left[\frac{G_1(r_+)+G_2(r_+)}{kr^3_+(r^2_++4\alpha)}+4\pi Pr^4_++3(r^2_++2\alpha)\right],
\end{equation}
where
\begin{eqnarray}
G_1(r_+)&=&\frac{q^2r_+}{3}\left[2k(r^2_++12\alpha)-3r^2_+(r^2_++4\alpha)\right]e^{-\frac{k}{r^2_+}},\nonumber\\
G_2(r_+)&=&q^2r^3_+(r^2_++4\alpha)-\frac{2kr^5_+}{3}(8\pi Pr^2_++3)(r^2_++12\alpha).
\end{eqnarray}
The behavior of the Gibbs free energy $G$ in terms of the temperature for various values of the pressure and the charge $q$ is plotted in Fig. \ref{Gibbs-energy}.
The Gibbs free energy $G$ changes its sign at which the Hawking-Page (HP) phase transition between the black hole and the thermal AdS space occurs \cite{Hawking1983}. We give the numerical results for the radius $r_{HP}$ and the temperature $T_{HP}$ for the Hawking-Page phase transition in Table \ref{HP-transition}.
\begin{table}[!htp]
\centering
\begin{tabular}{|c|c|c|c|c|c||c|c|c|c|c|}
  \hline
  & \multicolumn{5}{|c|}{$k=0.4$} & \multicolumn{5}{|c|}{$k=0.8$} \\
  \hline
  $P$ \ \ \ \ & 0.002 & 0.005 & 0.010 & 0.015 & 0.020 & 0.003 & 0.008 & 0.012 & 0.018 & 0.025 \\
  \hline
  $r_{HP}$ & 9.89057 & 4.97141 & 2.07537 & 1.76487 & 1.62517 & 7.59375 & 2.43735 & 1.88212 & 1.64399 & 1.50632 \\
  \hline
  $T_{HP}$ & 0.04161 & 0.06025 & 0.06955 & 0.0731 & 0.07582 & 0.04961 & 0.05534 & 0.071 & 0.07454 & 0.07026 \\
  \hline
  \end{tabular}
\caption{The numerical results for the radius $r_{HP}$ and the temperature $T_{HP}$ at the Hawking-Page phase transition, under the different values of the pressure and the nonlinear parameter, at $\alpha=0.5$ and $q=\sqrt{2}$.} \label{HP-transition}
\end{table}
From this table, one may see that the increase of the pressure or cosmological constant results in the decrease of the radius of the HP phase transition and the increase of the associated  temperature. 

In particular, there are swallowtail structures meaning that the black hole undergoes the first-order phase transition between the thermodynamically stable phases. More specifically, for the black holes belonging in the cases II and III, there is a first-order phase transition which disappears if the pressure is larger than the critical one $P_c$. On the other hand, for the black hole belonging in the case I, there exists, at least, one first-order phase transition point, irrespective of the pressure.

\section{\label{PV-crt} P-V criticality and Critical Exponents}
\begin{figure}[t]
 \centering
\begin{tabular}{cc}
\includegraphics[width=0.45 \textwidth]{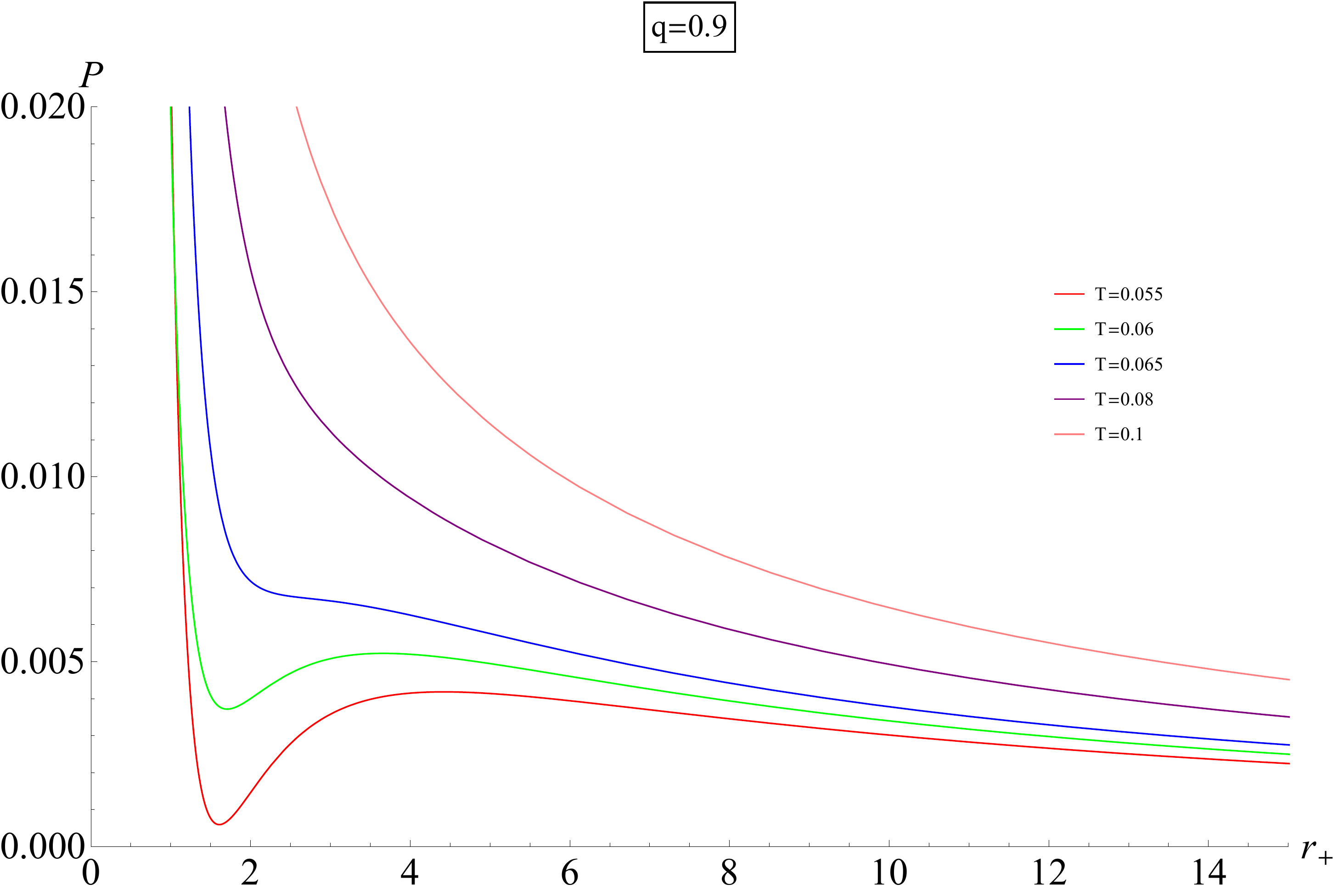}
\hspace*{0.05\textwidth}
\includegraphics[width=0.45 \textwidth]{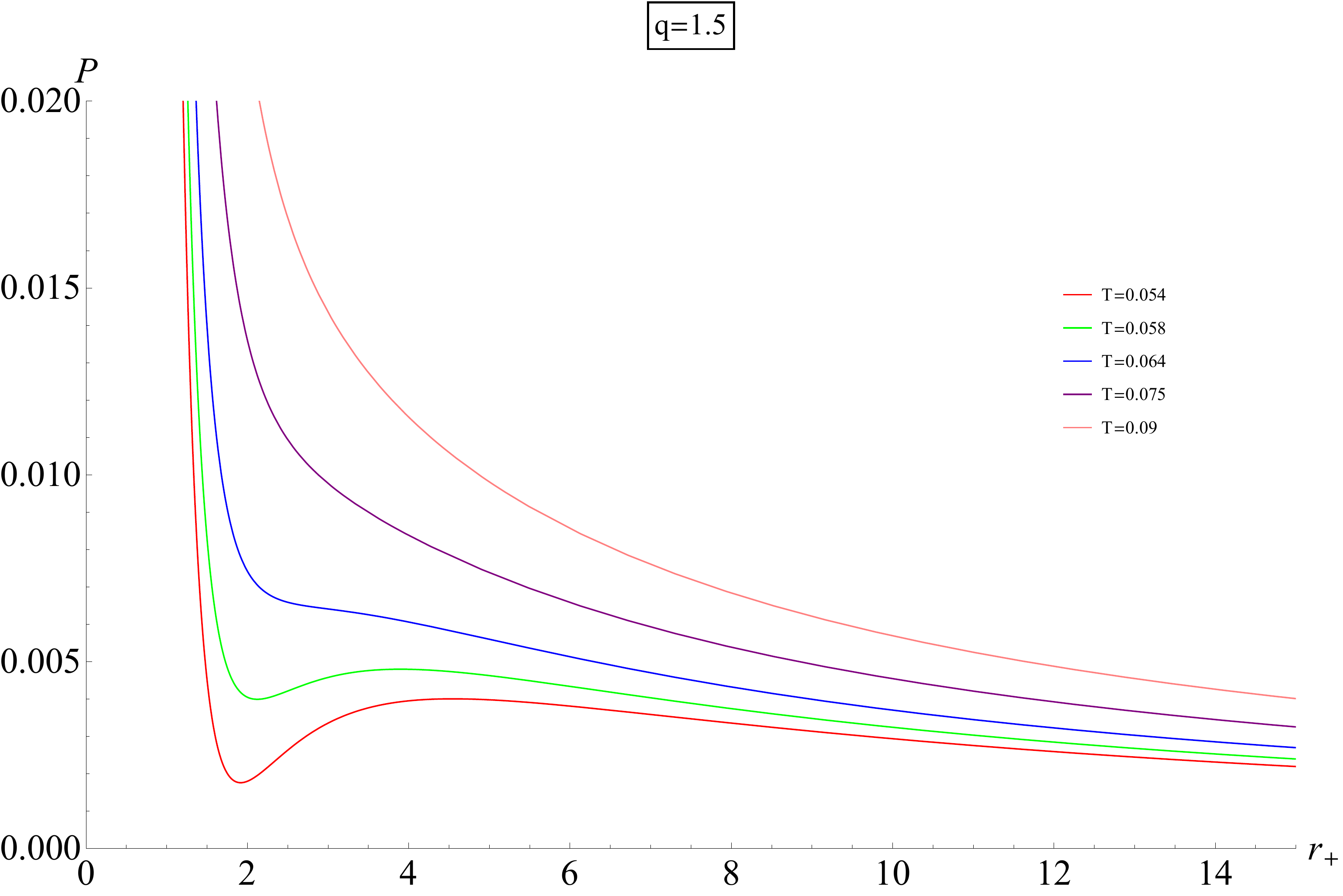}
\end{tabular}
  \caption{The isotherms in the $P-r_+$ diagram at $\alpha=0.5$ and $k=0.4$.}\label{PV-cr}
\end{figure}

In this section, we study the P-V criticality and critical exponents of the black holes in the extended phase space. From Eq. (\ref{bhtemp}), the equation of state $P=P(T,V)$ is given by
\begin{equation}
  P=\frac{3(r^2_++4\alpha)}{4r^3_+}T-\frac{3}{8\pi r^2_+}+\frac{q^2}{8\pi r^6_+}e^{-\frac{k}{r^2_+}},\label{bheqst}
\end{equation}
where $r_+$ is a function of the volume $V$, $r_{+}=\left(\frac{2}{\pi^{2}}V\right)^{1/4}$, from Eq. (\ref{thervol}). By comparing with the Van der Waals equation \cite{Mann2012}, one may identify the specific volume $v$ of the fluid with the event horizon radius $r_+$ as
\begin{equation}
v=\frac{16}{3}\frac{V}{\cal{A}}=\frac{4}{3}r_+,
\end{equation}
where $\cal{A}$ is the area of event horizon.
We plot the isotherms in the $P-r_+$ diagram in Fig. \ref{PV-cr}, which shows  
the various behavior of the isotherms depending on the temperature.

These isotherm curves show that the  system is in the 'ideal gas phase' for the temperature above the critical temperature $T_c$, at which 
 the isotherm has an inflection point,
\begin{equation}
\left(\frac{\partial P}{\partial
r_+}\right)_T=\left(\frac{\partial^2P}{\partial r^2_+}\right)_T=0,
\end{equation}
and has the 'liquid-gas phase transition'  for the temperature below the critical temperature $T_c$.

In $T<T_{c}$, the isotherms consist of three different branches of the black holes. The branch with very large $|\partial P/\partial r_+|$ or the high pressure is related to the small black holes. The branch with small $|\partial P/\partial r_+|$ or the low pressure corresponds to the large black holes. The oscillating branch between the small and  large black holes  is the region where the black holes are thermodynamically unstable and represents  a Van der Waals-like first-order phase transition between the small and large black holes. The oscillating part should be replaced by the straight line on which the pressure is constant under the change of the volume, or radius $r_{+}$. This phase transition pressure can be determined by Maxwell's equal area law, which will be given in below. The corresponding radius $r_c$ of the event horizon to the critical temperature is determined as
\begin{equation}
q^2\left[2k^2(r^2_c+12\alpha)+3r^4_c(5r^2_c+36\alpha)-kr^2_c(13r^2_c+132\alpha)\right]-3r^8_c(r^2_c-12\alpha)e^{\frac{k}{r^2_c}}=0.
\label{r_{c}}
\end{equation}
The critical temperature $T_c$ and the critical pressure are expressed in terms of  the critical radius $r_c$ as
\begin{eqnarray}
T_c&=&\frac{3r^6_c-q^2(3r^2_c-k)e^{-\frac{k}{r^2_c}}}{3\pi r^5_c(r^2_c+12\alpha)},\nonumber\\
P_c&=&\frac{3r^6_c(r^2_c-4\alpha)-q^2\left[r^2_c(5r^2_c+12\alpha)-2k(r^2_c+4\alpha)\right]e^{-\frac{k}{r^2_c}}}{8\pi r^8_c(r^2_c+12\alpha)}.
\end{eqnarray}

The equation (\ref{r_{c}}) can be solved in the following limits as
\begin{eqnarray}
r_c&=&2\sqrt{3\alpha} \ \ \ \ \text{as} \ \ q\rightarrow0,\nonumber\\
r_c&=&\left(5q^2\right)^{1/4} \ \ \ \ \text{as}\ \ \alpha,k\rightarrow0.
\end{eqnarray}
In the limit $q\rightarrow0$, the combination of these critical values becomes \cite{Cai2013}
\begin{equation}
\frac{P_cv_c}{T_c}=\frac{1}{3},
\end{equation}
whereas the limit $k,\alpha\rightarrow0$ leads to \cite{Gunasekaran2012,Chabab2012}
\begin{equation}
\frac{P_cv_c}{T_c}=\frac{5}{12}.
\end{equation}

Another interesting limit is the limit of strong nonlinear coupling, $k\gg 1$, which shows the effect of the nonlinear electrodynamics most clearly.
 In this limit, 
the exponential term in the Eq. (\ref{r_{c}}) is dominant and  the  critical radius becomes $r_c\simeq 2\sqrt{3\alpha}$. In this situation, the critical temperature and the critical pressure are given by
\begin{eqnarray}
T_c\simeq\frac{1}{4\pi\sqrt{3\alpha}},\qquad
P_c\simeq\frac{1}{96\pi\alpha}.
\end{eqnarray} 
These critical values are nothing but those corresponding to the case of the $q\rightarrow0$ limit. This seems to be natural since the sufficiently strong nonlinear electrodynamics means the strong repulsive force and thus the black holes tend to form with less charge.

In order to see more explicitly the effect of the nonlinear electrodynamics on $P-V$ criticality, we present the relation between the critical quantities and the nonlinear coupling, given numerically in Table. \ref{PV-sols}.
\begin{table}[!htp]
\centering
\begin{tabular}{|c|c|c|c|c|c|c|c|}
  \hline
  \multicolumn{4}{|c|}{$q=0.9$} & \multicolumn{4}{|c|}{$q=1.5$} \\
  \hline
  $k$ & $r_c$ & $T_c$ & $P_c$ & $k$ & $r_c$ & $T_c$ & $P_c$ \\
  \hline
  0.0 &  2.42  &  0.0706  &  0.00763  &  0.0  &  2.65  &  0.0681  &  0.00692  \\ \hline
  0.5 &  2.39  &  0.0708  &  0.00769  &  0.7  &  2.60  &  0.0685  &  0.00705  \\ \hline
  1.0 &  2.36  &  0.0710  &  0.00775  &  1.4  &  2.53  &  0.0670  &  0.00718  \\ \hline
  1.5 &  2.33  &  0.0711  &  0.00781  &  2.1  &  2.47  &  0.0694  &  0.00731  \\ \hline
  2.0 &  2.31  &  0.0713  &  0.00787  &  2.8  &  2.41  &  0.0698  &  0.00744  \\ \hline
  2.5 &  2.28  &  0.0715  &  0.00792  &  3.5  &  2.35  &  0.0702  &  0.00757 \\ \hline
  3.0 &  2.26  &  0.0716  &  0.00797  &  4.2  &  2.30  &  0.0706  &  0.00770  \\ \hline
  3.5 & 2.25   &  0.0717  &  0.00801  &  4.9  &  2.25  &  0.0710  &  0.00782  \\ \hline
  4.0 & 2.23   &  0.0719  &  0.00806  &  5.6  &  2.21  &  0.0713  &  0.00793  \\ \hline
  \end{tabular}
\caption{The numerical results for the critical radius, temperature and pressure are given in terms of the nonlinear coupling under various values of the electric charge, at $\alpha=0.4$.}\label{PV-sols}
\end{table}
As the nonlinear coupling increases, both the critical temperature and the critical pressure increase and approach eventually at the values $r_c=2.19$, $T_c=0.0726$, and $P_c=0.00829$ for $\alpha=0.4$, which correspond to the $q\rightarrow0$ (or $k\rightarrow\infty$) limit.

The behavior of the thermodynamic quantities near the critical point is described by the critical exponents which are defined as
\begin{eqnarray}
  C_V&=&T\frac{\partial S}{\partial T}\Big|_V\propto|t|^{-\alpha},\nonumber\\
  \eta&=&V_l-V_s\propto|t|^\beta,\nonumber\\
  \kappa_T&=&-\frac{1}{V}\frac{\partial V}{\partial P}\Big|_T\propto|t|^{-\gamma},\nonumber\\
  |P-P_c|\Big|_{T_c}&\propto&|V-V_c|^\delta,
\end{eqnarray}
where $t\equiv (T-T_c)/T_c$. The critical exponents $\alpha$, $\gamma$, and $\delta$ govern the
behavior of the specific heat at constant volume,   the isothermal compressibility $\kappa_{T}$, and  the critical isotherm, respectively.  The order parameter $\eta$  is given by  the difference of the volume $V_l-V_s$ during first order phase transition and its  critical exponent is denoted as   $\beta$.
The eqs. (\ref{bhent}) and (\ref{thervol}) imply that the black hole entropy $S$ is a function of the volume $V$ only, independent of the temperature $T$. This means that $C_V=0$ and thus  the critical exponent $\alpha=0$.

Now we would like to calculate remaining critical exponents following the approach given in Ref. \cite{Mann2012}. First, we define the dimensionless quantities near the critical point as
\begin{equation}
p=\frac{P}{P_c},\ \ \ \ 1+\epsilon=\frac{r_+}{r_c},\ \ \ \ 1+\omega=\frac{V}{V_c},\ \ \ \ 1+t=\frac{T}{T_c},
\end{equation}
with $|\epsilon|,|\omega|,|t|\ll1$. Note that, by replacing the above expansion of $r_+$ into Eq. (\ref{thervol}),  $\omega$ is related to $\epsilon$ as
$
\omega=4\epsilon
$.
Then, the equation of state, (\ref{bheqst}), is rewritten in terms of these dimensionless quantities as
\begin{equation}
  p=1+At-Bt\epsilon-C\epsilon^3+\mathcal{O}(t\epsilon^2,\epsilon^4),\label{press-exp}
\end{equation}
where
\begin{eqnarray}
  A&=&\frac{3(r^2_c+4\alpha)}{4r^3_c}\frac{T_c}{P_c},\nonumber\\
  B&=&\frac{3(r^2_c+12\alpha)}{4r^3_c}\frac{T_c}{P_c},\nonumber\\
  C&=&\frac{1}{P_c}\left[\frac{3(r^2_c+40\alpha)}{4r^3_c}T_c-\frac{3}{2\pi r^2_c}+\frac{q^2(84r^6_c-96kr^4_c+27k^2r^2_c-2k^3)}{12\pi r^{12}_c}e^{-\frac{k}{r^2_c}}\right].
\end{eqnarray}
We consider the case $t<0$.  Since the pressure is constant during the phase transition between the large and small black holes,  it leads to
\begin{equation}
1+At-Bt\epsilon_l-C\epsilon^3_l=1+At-Bt\epsilon_s-C\epsilon^3_s,\label{const-press}
\end{equation}
where $\epsilon_l$ and $\epsilon_s$ correspond to the radii of the large and small black holes, respectively. On the other hand, from Eq. (\ref{press-exp}), we can get the differentiation of the pressure with respect to the radius as
\begin{equation}
dP=-P_c(Bt+3C\epsilon^2)d\epsilon.
\end{equation}
Applying Maxwell's equal area law, we obtain
\begin{equation}
0=\int_{\omega_l}^{\omega_s}\omega dP=-4P_c\int_{\epsilon_l}^{\epsilon_s}\epsilon(Bt+3C\epsilon^2)d\epsilon.\label{Max-el}
\end{equation}
From Eqs. (\ref{const-press}) and (\ref{Max-el}), we have unique solution as
\begin{equation}
\epsilon_l=-\epsilon_s=\sqrt{\frac{-Bt}{C}}.
\end{equation}
Thus, we can determine the critical exponent $\beta$ as
\begin{equation}
   \eta=V_l-V_s=V_c(\omega_l-\omega_s)\propto\sqrt{-t}\ \ \Rightarrow\ \  \beta=\frac{1}{2}.
\end{equation}
In order to compute the critical exponent $\gamma$, let us calculate the isothermal compressibility
\begin{equation}
\kappa_T=-\frac{1}{V}\frac{\partial V}{\partial P}\Big|_T=\frac{4}{P_c(1+\omega)}\left(\frac{\partial p}{\partial\epsilon}\Big|_t\right)^{-1}\propto\frac{1}{Bt} \Rightarrow\ \  \gamma=1.
\end{equation}
Finally, we find the critical exponent $\delta$ by computing $|P-P_c|\Big|_{T_c}$. The critical isotherm corresponds to $T=T_c$ or $t=0$, hence we have
\begin{equation}
P\Big|_{T_c}=P_c(1-C\epsilon^3).
\end{equation}
As a result, the critical exponent $\delta$ is determined as
\begin{equation}
|P-P_c|\Big|_{T_c}\propto|\epsilon|^3\propto|V-V_c|^3\ \ \Rightarrow\ \ \delta=3.
\end{equation}
Quite surprisingly, all these critical exponents for the black holes with nonlinear electrodynamics have the same values as those of the Van der Waals fluid as well as the black holes with the Maxwell electrodynamics.

\section{\label{conclu} Conclusion}
One of the main motivation of this work has been to find the regular black hole solution which does not have the curvature singularity. One way to obtain such a solution is to smooth out the geometry by adding nontrivial source like $U(1)$ gauge field with higher derivatives. In this work, we have found a new class of $5D$  charged AdS black hole solutions in Einstein-Gauss-Bonnet gravity coupled to a nonlinear
electromagnetic field. In particular, the solutions admit not only extremal black holes, but also regular black hole solution. 

In general, the regular black hole has non-zero Hawking temperature and thus radiates. Our solution  suggests various interesting scenarios on the fate of the black holes after Hawking radiation. Depending on the initial state of black holes, the end point of our black holes would be the extremal black hole with curvature singularity or would be the  regular extremal black hole. The perspective to obtain extremal black holes without any curvature singularity, even with small possibility, is very exciting. 

We have investigated the thermodynamics and the phase transitions of these black holes, in the extended phase space where the cosmological constant is taken into account as the pressure.  Based on the behavior of heat capacity at constant pressure, we have found the second-order phase transitions. And by studying the behavior of the Gibbs free energy, we have found the Hawking-Page phase transition. Through the $P-r_+$ diagram we have shown that the black hole can undergo liquid/gas-like phase transition if the black hole temperature is below a critical value. Finally, we have computed the critical exponents which govern the behavior of the specific heat at constant volume, the order parameter, the isothermal compressibility and the critical isotherm near the critical point and found that they have the same values as those for the Van der Waals fluid.

\section*{Acknowledgements}
We would like to thank Dr. Kyung Kiu Kim for useful discussion. This work was supported by the National Research Foundation of Korea(NRF) grant with the grant number NRF-2016R1D1A1A09917598 and by the Yonsei University Future-leading Research Initiative of 2017(2017-22-0098).

\section*{Appendix I}
In this appendix, we give a detailed EOM and the computation of the black hole entropy through Wald formula. 
For the configuration, 
\begin{eqnarray}
ds^2&=&-f(r)dt^2+f(r)^{-1}dr^2+r^2d\Omega^2_3,\nonumber\\
P_{tr}&=&\frac{q}{r^{3}}.
\end{eqnarray}
the non-zero components  of Eq. (\ref{Eeq}) are given by
\begin{eqnarray}
\bar{G}^t_t&=&\bar{G}^r_r=-\frac{6}{l^2}-\frac{6\alpha[f(r)-1]f'(r)}{r^3}+\frac{3[-2+2f(r)+rf'(r)]}{2r^2},\nonumber\\
\bar{G}^\psi_\psi &=&\bar{G}^\theta_\theta=\bar{G}^\phi_\phi=-\frac{6}{l^2}+\frac{-1+f[r]+2rf'(r)+r^2f''(r)/2}{r^2}-\frac{2\alpha\{f'(r)^2+[f(r)-1]f''(r)\}}{r^2},\nonumber\\
\bar{T}^t_t&=&\bar{T}^r_r=-\frac{q^2}{r^6}e^{-\frac{k}{r^2}},\nonumber\\
\bar{T}^\psi_\psi&=&\bar{T}^\theta_\theta=\bar{T}^\phi_\phi=\frac{q^2(3r^2-2k)}{3r^8}e^{-\frac{k}{r^2}},
\end{eqnarray}
where we have denoted
\begin{eqnarray}
\bar{G}_\mu^\nu &\equiv&G^\nu_\mu+\alpha H^\nu_\mu-\frac{6}{l^2}\delta^\nu_\mu,\nonumber\\
\bar{T}_\mu^\nu  &\equiv&2\left[\frac{\partial\mathcal{H}}{\partial
P}P_{\mu\rho}P^{\nu\rho}-\delta^\nu_\mu\left(2\frac{\partial\mathcal{H}}{\partial P}P-\mathcal{H}\right)\right].
\end{eqnarray}
It is easily to see that the $(\psi\psi)$, $(\theta\theta)$ and $(\phi\phi)$ components of Eq. (\ref{Eeq}) are the consequence of  $(tt)$ or $(rr)$ component of Eq. (\ref{Eeq}). The $(tt)$ component of Eq. (\ref{Eeq}) reads
\begin{eqnarray}
\left[4\alpha f'(r)-2r\right]\left[f(r)-1\right]-r^2f'(r)+\frac{4r^3}{l^2}=\frac{2q^2}{3r^3}e^{-\frac{k}{r^2}}.
\end{eqnarray}
The curvature scalars $R$, $R_{\mu\nu}R^{\mu\nu}$, and $R_{\mu\nu\rho\lambda}R^{\mu\nu\rho\lambda}$ become
\begin{eqnarray}
R&=&-\frac{1}{r^2}\left[r^2f''(r)+6rf'(r)+6f(r)-6\right],\nonumber\\
R_{\mu\nu}R^{\mu\nu}&=&\frac{3}{r^4}\left[rf'(r)+2f(r)-2\right]^2+\frac{1}{2r^2}\left[rf''(r)+3f'(r)\right]^2,\nonumber\\
R_{\mu\nu\rho\lambda}R^{\mu\nu\rho\lambda}&=&\frac{1}{r^4}\left\{r^4\left[f''(r)\right]^2+6r^2\left[f'(r)\right]^2+12\left[f(r)\right]^2-24f(r)+12\right\}.
\end{eqnarray}

The black hole entropy is given by the Wald formula, which is defined by
\begin{eqnarray}
S&=&-2\pi\oint_{\Sigma}\frac{\delta\mathcal{L}}{\delta R_{\mu\nu\rho\sigma}}\epsilon_{\mu\nu}\epsilon_{\rho\sigma} dV_3,\nonumber\\
&=&-\frac{1}{8}\oint_{\Sigma}\left[\frac{\delta R}{\delta R_{\mu\nu\rho\sigma}}+\alpha\frac{\delta\mathcal{L}_{GB}}{\delta R_{\mu\nu\rho\sigma}}\right]\epsilon_{\mu\nu}\epsilon_{\rho\sigma}dV_3,
\end{eqnarray}
where the binormal vector $\epsilon_{\mu\nu}$ to the event horizon surface $\Sigma$ and the volume element on $\Sigma$ are given by
\begin{equation}
\epsilon_{\mu\nu}=\delta^t_\mu\delta^r_\nu-\delta^r_\mu\delta^t_\nu,\ \ \ \ dV_3=r^3_+d\Omega_3.
\end{equation}
 With the metric ansatz given in Eq. (\ref{charge-P}), we have
\begin{eqnarray}
\frac{\delta R}{\delta R_{\mu\nu\rho\sigma}}\epsilon_{\mu\nu}\epsilon_{\rho\sigma}&=&g^{\mu\rho}g^{\nu\sigma}\epsilon_{\mu\nu}\epsilon_{\rho\sigma}=-2,\nonumber\\
\frac{\delta\mathcal{L}_{GB}}{\delta R_{\mu\nu\rho\sigma}}\epsilon_{\mu\nu}\epsilon_{\rho\sigma}&=&-4R-8\left(R^{tt}g^{rr}+R^{rr}g^{tt}\right)+2\left(R^{trtr}-R^{rttr}-R^{trrt}+R^{rtrt}\right)=-\frac{24}{r^2_+}.
\end{eqnarray}
As a result, we find the black hole entropy as
\begin{equation}
S=\frac{S_3r^3_+}{4}\left(1+\frac{12}{r^2_+}\right).
\end{equation}

\section*{Appendix II}
In this appendix, we describe the deflection of the photon from the black holes. It should be stressed that the influence of the nonlinear electrodynamics on the motion of the photon in the charged black hole background has been investigated in various works. In such a situation, the motion of the photon is not just given by the null geodesic of the spacetime geometry, but the null geodesics of the effective geometry, which reflects the nonlinear feature of electrodynamics \cite{Novello2000,Novello2001,Obukhov2002,Bergliaffa2009,JSchee2015a,JSchee2015b,JSchee2016}. This could be confirmed by studying the electromagnetic perturbations in the eikonal regime \cite{Toshmatov2019}. As a result, it was demonstrated in the regular Bardeen black holes that the nonlinear electrodynamics has the strong influence on the deflection angle of the photon and black hole shadows \cite{Schee2019a} as well as in the image of Keplerian disks \cite{Schee2019b}. Here, we want to focus on the corrections from the Gauss-Bonnet term on the behavior of the photon in the black hole spacetime, and thus take the Maxwell limit, $k\rightarrow 0$. Furthermore, we restrict ourselves to the case of weak deflection only, while leaving the full investigation of trajectories of the photon, including the effect of the nonlinear electrodynamics, for the future.

The Lagrangian describing the null geodesics with the metric (\ref{fr-funct}) is given by
\begin{eqnarray}
2\mathcal{L}=g_{\mu\nu}\frac{dx^\mu}{d\lambda}\frac{dx^\nu}{d\lambda}=-f(r)\dot{t}^2+\frac{\dot{r}^2}{f(r)}+r^2\left[\dot{\theta}^2+\sin^2\theta\left(\dot{\psi}^2+\sin^2\psi\dot{\phi}^2\right)\right],
\end{eqnarray}
where $\lambda$ is the affine parameter and the dots denote the derivative with respect to $\lambda$. Because the Lagrangian is independent of  coordinates $t$ and $\phi$, the corresponding conjugate momenta are conserved as
\begin{eqnarray}
\frac{\partial\mathcal{L}}{\partial\dot{t}}&=&-f(r)\dot{t}=-E,\nonumber\\
\frac{\partial\mathcal{L}}{\partial\dot{\phi}}&=&r^2\sin^2\theta\sin^2\psi\dot{\phi}=L,
\end{eqnarray}
where $E$ and $L$ are integration constants corresponding to the energy and angular momentum of the photon, respectively. The Euler-Lagrange equations for the coordinates $\theta$ and $\psi$ are given by
\begin{eqnarray}
\frac{d}{d\lambda}\left(r^2\dot{\theta}\right)&=&r^2\sin\theta\cos\theta\left(\dot{\psi}^2+\sin^2\psi\dot{\phi}\right),\nonumber\\
\frac{d}{d\lambda}\left(r^2\dot{\psi}\right)&=&r^2\sin\psi\cos\psi\sin^2\theta\dot{\phi}.
\end{eqnarray}

Without loss of generality, we can simplify the situation by choosing the initial conditions as $\dot{\theta}=\dot{\psi}=0$, $\theta=\psi=\frac{\pi}{2}$. For the geodesic of the (incoming) photon, we have $\mathcal{L}=0$, which leads to the following equation
\begin{eqnarray}
\frac{d\phi}{du}=\left[\frac{1}{b^2}-u^2f(u)\right]^{-1/2},\label{phi-u-equ}
\end{eqnarray}
where $b=L/E$, $u=1/r$ and
\begin{equation}
f(u)=1+\frac{1}{4\alpha u^2}\left(1-\sqrt{1-\frac{8\alpha}{l^2}+8\alpha u^4\left(m+\frac{q^{2}}{3k}(e^{-ku^2}-1)\right)}\right).
\end{equation}
From now on, we take the Maxwell limit, $k\rightarrow 0$, to focus on the corrections from the Gauss-Bonnet term. By following \cite{BShutz2009} one can write Eq. (\ref{phi-u-equ}) in the weak deflection limit as
\begin{eqnarray}
\frac{d\phi}{dy}=\frac{6+9\bar{m}y^2-5\bar{q}^2y^4}{6\sqrt{\bar{b}^{-2}-y^2}}+\mathcal{O}(m^2u^4,q^4u^8),
\end{eqnarray}
where
\begin{eqnarray}
y\equiv u\left(1-\bar{m}u^2+\frac{\bar{q}^2u^4}{3}\right)^{\frac{1}{2}}\ \ \rightarrow \ \ u\simeq y\left(1+\frac{\bar{m}y^2}{2}-\frac{\bar{q}^2y^4}{6}\right),
\end{eqnarray}
and
\begin{eqnarray}
\frac{1}{\bar{b}^2}\equiv\frac{1}{b^2}-\frac{1}{l^2_{\text{eff}}}, \ \ \ \ \bar{m}\equiv\left(1-\frac{4\alpha}{l^2_{\text{eff}}}\right)^{-1}m, \ \ \ \ \bar{q}\equiv\left(1-\frac{4\alpha}{l^2_{\text{eff}}}\right)^{-1/2}q.
\end{eqnarray}
Integrating this equation leads to
\begin{eqnarray}
\phi(y)&=&\frac{y\left[15\bar{q}^2+\bar{b}^2\left(10\bar{q}^2y^2-36\bar{m}\right)\right]}{48\bar{b}^2}\sqrt{\frac{1}{\bar{b}^2}-y^2}+\frac{16\bar{b}^4+12\bar{b}^2\bar{m}-5\bar{q}^2}{16\bar{b}^4}\text{arcsin}(\bar{b}y)+\phi_0,\nonumber\\
\end{eqnarray}
where $\phi_0$ is the incoming angle of the photon. As a result, one can find the deflection angle of the photon as
\begin{eqnarray}
\Delta\phi&=&2\left[\phi(1/\bar{b})-\phi_0\right]-\pi,\nonumber\\
&=&\Delta\phi_{\text{GR}}+\pi\left[\frac{3(2l^2-3b^2)}{2b^2l^4}m-\frac{5(2b^4-3b^2l^2+l^4)}{4b^4l^6}q^2\right]\alpha+\mathcal{O}(\alpha^2),\label{defang}
\end{eqnarray}
where $1/\bar{b}$ corresponds to the nearest point of the photon's trajectory and $\Delta\phi_{\text{GR}}$ is the standard value predicted by GR, given as
\begin{eqnarray}
\Delta\phi_{\text{GR}}=\pi\left[\frac{3(l^2-b^2)}{4b^2l^2}m-\frac{5(l^2-b^2)^2}{16b^4l^4}q^2\right].
\end{eqnarray}
The second term in (\ref{defang}) shows the correction coming from the Gauss-Bonnet term to GR.

\end{document}